\documentclass[%
 reprint,pre,doublecol,showkeys,showpacs
]{revtex4-2}
\usepackage{hyperref}
\hypersetup{
  colorlinks   = true,    % Colours links instead of ugly boxes
  urlcolor     = blue,    % Colour for external hyperlinks
  linkcolor    = blue,    % Colour of internal links
  citecolor    = red      % Colour of citations
}
\usepackage{graphicx}
\usepackage{lipsum}
\usepackage{romannum}
\usepackage{amsmath}
\usepackage{amssymb}
\usepackage{comment}
\usepackage{xcolor}
\usepackage{bm}% bold math
\begin{document}

\preprint{APS/123-QED}

\title{Stationary states of an active Brownian particle in a harmonic trap}

\author{Urvashi Nakul}
 %\altaffiliation[Also at ]{IIT Madras, Chennai.}%Lines break automatically or can be forced with \\
\author{Manoj Gopalakrishnan}%
% \email{urvashinakul123@gmail.com}
\affiliation{%
Department of Physics, Indian Institute of Technology Madras, Chennai 600036, India\\
}%
\date{\today}% It is always \today, today,

\begin{abstract}
We study the stationary states of an over-damped active Brownian particle (ABP) in a harmonic trap in two dimensions, via mathematical calculations and numerical simulations. In addition to translational diffusion, the ABP  self-propels with a certain velocity, whose magnitude is constant, but its direction is subject to Brownian rotation. In the limit where translational diffusion is negligible, the stationary distribution of the particle's position shows a transition between two different shapes, one with maximum and the other with minimum density at the centre, as the trap stiffness is increased. We show that this non-intuitive behaviour is captured by the relevant Fokker-Planck equation, which, under minimal assumptions, predicts a continuous phase transition-like change between the two different shapes. As the translational diffusion coefficient is increased, both these distributions converge into the equilibrium, Boltzmann form. Our simulations support the analytical predictions, and also show that the probability distribution of the orientation angle of the self-propulsion velocity undergoes a transition from unimodal to bimodal forms in this limit. We also extend our simulations to a three dimensional trap and find similar behavior. 
\end{abstract}


\maketitle

%\tableofcontents

\section{\label{sec:level1}INTRODUCTION}

The study of self-propelled Brownian particles is in great demand in physical and biological research communities~\cite{marchetti2013hydrodynamics,ramaswamy2017active}. These active particles are capable of converting energy from the environment into directional movement~\cite{schweitzer2003brownian}. Active or self-propelled particle model describes a wide range of natural and artificial systems, e.g., vibrated granular matter~\cite{walsh2017noise}, cellular cytoskeleton~\cite{ramaswamy2010mechanics}, flocking of birds~\cite{vicsek2012collective} etc.

In nature, self-propelled particles are frequently observed as suspensions in confined fluids. For instance, chromosomes are naturally present within the cell nuclei \cite{alberts2008molecular}. The study of the dynamics of active particles in enclosed environments is crucial for understanding their behavior in biological systems. Several experimental techniques have been utilized to investigate the dynamics of these particles, including acoustic traps \cite{takatori2016acoustic}, active baths \cite{maggi2014generalized,pincce2016disorder}, and optical traps \cite{schmidt2018microscopic}. Furthermore, researchers have explored the impact of hydrodynamic interactions on the dynamics of self-propelled particles due to geometric boundaries. The investigation has shown that hydrodynamic coupling affects the dynamics of self-propelled microswimmers and leads to the emergence of non-intuitive behavior \cite{li2017two, debnath2018hydrodynamic, debnath2019active}. Notably, the dynamics of individual particles also exhibit intriguing phenomena, such as accumulation near confined boundaries \cite{li2009accumulation,wagner2017steady} and non-Boltzmann distribution in stationary states \cite{tailleur2008statistical,maggi2014generalized,das2018confined}. In this work, we aim to investigate the non-equilibrium stationary state of active particles confined in a harmonic trap.

Several models are in use to describe such motion, the most common being (i) Active Brownian particle (ABP), (ii) Active Ornstein-Uhlenbeck particle  (AOUP) and (iii) Run and Tumble particle (RTP). In the ABP model~\cite{schimansky1995structure,romanczuk2012active,fily2012athermal,bechinger2016active}, the particle self-propels with a constant speed $u_0$ in the heading direction characterized by unit vector $\hat{\bf u}$, and undergoes reorientation due to  thermal noise, which is uncorrelated and Gaussian. In AOUP model~\cite{szamel2014self,fodor2016far}, on the other hand, the propulsion speed is also variable and is Gaussian-distributed. Yet another common example of a self-propelled particle is a run and tumble particle (RTP)~\cite{schnitzer1993theory,tailleur2008statistical,solon2015active}, which undergoes abrupt and discontinuous changes in direction (tumbles), in addition to the continuous reorientation due to Brownian rotations. RTP is often used as a minimal model for swimming and tumbling bacteria like {\it Escherichia coli}~\cite{cates2013active,angelani2014first,angelani2017confined,dhar2019run}. Recently, another model of self-propelling Brownian particles, called the ``parental active model" (PAM)~\cite{caprini2022parental} has been proposed, which interpolates between AOUP and ABP models. In this paper, we focus only on stationary states of ABP in a harmonic trap. 

Understanding the non-equilibrium stationary states of confined active particles is crucial in developing our knowledge of the statistical physics of active matter. Over the years, several studies on self-propelled particles have highlighted various aspects of their non-equilibrium, and often counter-intuitive behaviour~\cite{maude1963non,berke2008hydrodynamic,nash2010run,yang2014aggregation,fily2014dynamics,wang2014diffusion}, some works related to unusual transport features of active particles, e.g., autonomous ratcheting \cite{ghosh2013self, reichhardt2017ratchet, bag2022directed},  Giant negative mobility and Enhanced buoyancy of active particles. 
In recent times, a number of experimental~\cite{li2009accumulation,maggi2014generalized,takatori2016acoustic,dauchot2019dynamics,schmidt2021non,buttinoni2022active} and theoretical~\cite{pototsky2012active,elgeti2015run,basu2018active,das2018confined,basu2019long,dhar2019run,malakar2020steady,chaudhuri2021active,santra2021direction,santra2021active} papers have reported significant developments in our understanding of the statistical physics of ABP. Takatori et al.~\cite{takatori2016acoustic} explored the dynamics of Janus particles in a  two-dimensional harmonic potential, by tuning the trap strength. They observed that the stationary state distribution of the particle's position changes from  Gaussian-like in weak trap to a strongly non-Gaussian, {\it active} distribution under strong trapping. Schmidt et al.~\cite{schmidt2021non} also observed  similar features in active nano-particles, which were confined in a potential well by varying the laser power. More recently, Buttinoni et al.~\cite{buttinoni2022active} studied active particles in harmonic potential and confirmed Boltzmann-like distribution in the absence of activity. In the presence of activity, at a certain value of trap strength the spatial distribution was found to become non-Gaussian. 

On the theoretical front, Basu et al.~\cite{basu2018active} showed that the short-time dynamics of ABP in two dimensions is strongly non-diffusive, with anomalous first passage properties. Malakar et al.~\cite{malakar2020steady} obtained an exact series solution to the underlying Fokker-Planck equation (FPE) for the stationary state positional distribution of the ABP in a two-dimensional trap, by expanding about the equilibrium (Boltzmann) distribution, which is Gaussian. As the trap stiffness $\kappa$ is increased, keeping other parameters fixed, their results suggested a transition from ``passive" to ``active" phases, and then a re-entrant transition to the ``passive" phase. Basu et al.~\cite{basu2019long} studied the stationary state of ABP in a two-dimensional harmonic potential well and derived exact expressions for the time-dependent moments of the particle's position. They showed that, for vanishing translational diffusion,  the stationary distribution becomes asymptotically Gaussian in the limit $\kappa\to 0$, whereas, in the opposite limit of very strong trapping, the distribution becomes ring-like. However, the Gaussian distribution here is characterised by an effective, ``active'' temperature, distinct from the temperature of the surrounding fluid medium. Chaudhuri and Dhar~\cite{chaudhuri2021active} extended the calculation of the moments to arbitrary dimensions, and constructed a kurtosis heat map of the system, analogous to a phase diagram, which clearly showed deviations from the Gaussian behaviour for the positional distribution. Both the above papers~\cite{basu2019long,chaudhuri2021active} show clearly that in the limit of very large translational diffusion coefficient $D_t$, the positional distribution asymptotically approaches the equilibrium, Boltzmann form, which is Gaussian. 

In spite of the above developments, the nature of the positional distribution of the ABP in the limit $D_t\to 0$ remains incompletely understood, in our opinion. It is, by now, well-established that this distribution changes from a Gaussian-like, concave-shaped (around the origin) form in the limit of small trap stiffness, to a strongly non-equilibrium, ring-like shape (in two dimensions) for very strong trapping. However, the nature of this transition remains ill-understood, and the mathematical form of the probability distribution which describes these active distributions is not obvious. With these gaps as our primary motivations, in this paper, we revisit the problem of ABP in an isotropic harmonic trap, with the goal of obtaining a deeper understanding of the non-trivial properties of the stationary states, and the transitions between them. 

We report significant analytic progress in obtaining a closed form solution of the Fokker-Planck equation for the position of the ABP, in the limit of small $D_t$. This is achieved in two steps: (a) by constructing an {\it exact} Fokker-Planck equation for the particle's position after integrating out the orientation angle of the propulsion velocity, and (b) estimating the radial dependence of the moments of the above angle using symmetry arguments as well as by using a Gaussian approximation for the corresponding angular distribution. The closed form solution thus obtained agrees well with numerical simulations. Our results, both analytical and numerical, strongly suggest a continuous phase transition-like change between two general shapes: (i) ``concave'' at the origin and (ii) ``convex'' at the origin, at a certain critical value of the trap stiffness. Furthermore, both the distributions have finite support and vanish identically for $r>r_m$, where $r$ is the radial distance from the trap's centre and $r_m$ is an intrinsic length scale. While both distributions are, in general, non-Gaussian, the concave distribution reduces to the expected ``active'' Gaussian form in the limit $\kappa\to 0$, in agreement with existing theoretical results~\cite{basu2019long}.  The angular distribution of ABP also undergoes a transition, from unimodal to bimodal, as the critical trap strength is exceeded. While the particles are predominantly oriented radially outwards from the trap's centre in weak (i.e., sub-critical) trap, they orbit around it for strong, super-critical trap strength. We also extend our simulations to three dimensions and observe a similar transition, but at a different critical value of the trap strength. For large $D_t$, our simulations reproduce the expected crossover to the equilibrium, Boltzmann form, as expected from general scaling arguments and physical intuition. 

The paper is organized as follows. In Sec.\ref{sec:level2}, we introduce and solve the general Fokker Planck Equation(FPE) for the dynamics of ABP confined under two-dimensional isotropic harmonic trap. 
In Sec.\ref{sec:level3}, we present and discuss the results from numerical simulations, for both two and three-dimensional traps, and include comparisons with our analytical predictions wherever applicable. We conclude by summarizing our principal results in Sec.\ref{sec:level4}. 

\section{\label{sec:level2}The Fokker-Planck equation and its solution}

Consider an ABP in a $d$-dimensional harmonic trap, with potential energy $U({\bf r})=(\kappa/2)r^2$, where $\kappa$ is the trap stiffness and ${\bf r}$ is the position vector of the particle measured with respect to the trap centre. Let ${\bf u}=u_0\hat{\bf u}$ be the instantaneous, active propulsion vector such that $u_0\equiv |{\bf u}|$ denotes the (constant) speed of propulsion. 
Let $\zeta_t^{-1}$ be the translational mobility of the particle in the viscous medium, and let $\zeta_r^{-1}$ be the rotational mobility. In the over-damped limit, the following Langevin equations describe the dynamics of the particle~\cite{peruani2010cluster,fily2012athermal,redner2013structure,wysocki2014cooperative,winkler2015virial,das2018confined}: 
\begin{equation}
    \begin{aligned}
        &\frac{d\mathbf{r}}{dt}=u_0\hat{\bf u}-k \textbf{r}+\sqrt{2D_t}{\boldsymbol \eta}_t(t),\\
        &\frac{d\hat{\bf u}}{dt}=\sqrt{2D_r}{{\boldsymbol \eta}_r(t)}\times \hat{\bf u}(t),
    \end{aligned} \label{eq:langevin_main}
\end{equation}
where $k=\kappa \zeta_t^{-1}$ and ${\bm \eta}_t$ and ${\bm \eta}_r$ are two Gaussian white noise terms with zero mean, whose different components are uncorrelated with each other, and $\langle{\bm \eta}_t(t)\cdot {\bm \eta}_t(t^{\prime})\rangle=d \,\delta(t-t^{\prime})$ and $ \langle{\bm \eta}_r(t)\cdot {\bm \eta}_r(t^{\prime})\rangle=(d-1)\, \delta(t-t^{\prime})$. In Eq.~\ref{eq:langevin_main}, $D_t=k_BT/\zeta_t$ and $D_r=k_BT/\zeta_r$ are the thermal translational and rotational diffusivities of the particle. 

The corresponding Fokker-Planck equation for the probability distribution $P(\textbf{r},\hat{\bf u},t)$ is given by
    \begin{align}
    \frac{\partial P}{\partial t}=&-\nabla \cdot \textbf{J}_t+ D_r \nabla^2_{\hat{\bf u}} P \label{eq:fpe}
    \end{align}
where  $\nabla^2_{\hat{\bf u}}$ is the angular Laplacian and 
\begin{equation}
{\bf J}_t=(-D_t\nabla + {\bf v})P
\label{eq:stationarycurrent}
\end{equation}
is the probability current density corresponding to translational motion. Here, 
\begin{equation}
{\bf v}=u_0\hat{\bf u}-k \textbf{r}
\label{eq:v}
\end{equation}
is the deterministic part of the velocity of the ABP. In the stationary state, the time derivative on the l.h.s of Eq.~\ref{eq:fpe} vanishes, which leads to the stationary state equation 
\begin{equation}
    \nabla\cdot {\bf J}_t=D_r\nabla^2_{\hat{\bf u}} P
    \label{eq:stationaryFPE}
\end{equation}
Let us define the marginal probability distribution for the particle's position 
\begin{equation}
          \Phi(\textbf{r})=\int P(\textbf{r},\hat{\bf u})~d^d\hat{\bf u}, 
        \label{eq:r4}
 \end{equation}
where $d^d\hat{\bf u}$ symbolically represents the angular integration in $d$ dimensions. After integrating Eq.~\ref{eq:stationaryFPE} over $\hat{\bf u}$, $\Phi({\bf r})$ is found to satisfy the equation 
 \begin{equation}
 \nabla\cdot {\bf K}=0, 
 \label{eq:new}
 \end{equation}
 where 
 \begin{equation}
     {\bf K}(\textbf{r})= -D_t\nabla\Phi+
     (u_0\overline{\hat{\bf u}}({\bf r})- k \textbf{r})\Phi(\textbf{r})  \label{eq:kr}
 \end{equation}
is the effective probability current for the position variable, and 
\begin{equation}
    \overline{\hat{\bf u}}=\Phi({\bf r})^{-1}\int \hat{\bf u}~P(\textbf{r},{\hat{\bf u}}) ~d^d\hat{\bf u}
\end{equation}
is the conditional expectation of the unit propulsion vector. 
From the symmetry of the potential, it follows that the tangential component $K_{\phi}=0$, which, when used in Eq.~\ref{eq:new} leads to the  conclusion that the radial component $K_r=0$ as well (since there are no sources or sinks for particles anywhere). We thus conclude that ${\bf K}({\bf r})=0$ identically. The radial symmetry of the potential also implies that  
\begin{equation}
   \overline{\hat{\bf u}}({\bf r})=\lambda(r)\hat{\bf r} 
   \label{eq:gr}
\end{equation}
for some function $\lambda(r)\leq 1$, while $\Phi({\bf r})=\Phi(r)$.With the l.h.s thus put to zero, Eq.~\ref{eq:kr} is conveniently expressed in the following dimensionless form, in terms of a dimensionless position vector ${\boldsymbol \xi}={\bf r}/r_m$, where 
\begin{equation}
    r_m=u_0/k
    \label{eq:rm}
\end{equation}
is an important length scale in the problem. With the l.h.s put to zero in stationary state, Eq.~\ref{eq:kr} becomes, in dimensionless form, 
\begin{equation}
    -D_t^{\prime}\frac{\partial \Psi}{\partial \xi} +(g(\xi)-\xi)\Psi=0
    \label{eq:dimensionless}
\end{equation}
where $g(\xi)=\lambda(r_m\xi)$, $\Psi(\xi)=r_m^d\Phi(r_m\xi)$, and  
\begin{equation}
    D_t^{\prime}=D_t k/u_0^2. 
    \label{eq:dimensionlessparameter}
\end{equation}
is a dimensionless diffusion coefficient, which characterizes the relative importance of translational diffusion in the problem, in comparison with activity. Also, when expressed in terms of $r_m$, $D_t^{\prime}=D_t/r_m u_0$  takes the form of an inverse P\'{e}clet number.

The formal solution to Eq.~\ref{eq:dimensionless} is given by 
\begin{equation}
\Psi(\xi)=\Psi(0) \exp\bigg(-\frac{1}{D_t^{\prime}}\int_{0}^{\xi}[\xi^{\prime}-g(\xi^{\prime})]d\xi^{\prime}\bigg), 
\label{eq:psi}
\end{equation}
which is essentially the equilibrium Boltzmann distribution, modified by activity. In fact, after using the Einstein relation $D_t=k_B T/\zeta_t$, Eq.~\ref{eq:psi} can be recast in the equivalent form 
\begin{equation}
    \Phi(r)=\Phi(0)e^{-\frac{\kappa r^2}{2k_B T}} \exp{\bigg(\frac{1}{D_t^{\prime}}\int_{0}^{r/r_m} g(r^{\prime}) dr^{\prime}\bigg)}. 
    \label{eq:formalsolution}
\end{equation}
The first exponential term is independent of the active velocity $u_0$, while the second one depends on it. As the active velocity $u_0$ is increased, keeping other parameters fixed, the second term gains in significance and the probability distribution deviates significantly from the equilibrium Gaussian form. In the following section, we derive a closed mathematical form for this non-equilibrium distribution, under certain assumptions. In the limit 
$D_t^{\prime}\to 0$, it also turns out that $r_m$ is the maximum radial distance achieved by the ABP. 

\subsection{The strongly active regime: $D_t^{\prime}\to 0$}
In the limit $D_t^{\prime} \ll 1$, $u_0\gg \sqrt{kD_t}$ and active propulsion dominates over translational diffusion. In this limit, the gradient term in Eq.~\ref{eq:dimensionless} can be neglected in comparison with the second term, which leads to the relation 
\begin{equation}
g(\xi)\simeq \xi~~~(D_t^{\prime}\to 0).
\label{eq:gr1}
\end{equation}
In terms of the original variables, the equivalent relation is 
\begin{equation}
     \overline{\hat{\mathbf{u}}}(\mathbf{r})\simeq \frac{\mathbf{r}}{r_m}~~~~~~(D^{\prime}_t\to 0).
     \label{eq:u_bar}
  \end{equation}
We now show that, using the relation in Eq.~\ref{eq:u_bar}, a closed mathematical form for $\Phi(r)$ can be derived in this limit, under minimal  assumptions. Note that, in this activity-dominated limit, the solutions are expected to be strongly non-equilibrium, stationary state  distributions. 

In order to solve for the distribution $\Phi(r)$, let us multiply Eq. \ref{eq:stationaryFPE} by the velocity $\textbf{v}$ in Eq.~\ref{eq:v} and integrate over $\hat{\textbf u}$, which leads to the equation 
\begin{figure}
\includegraphics[width=\linewidth]{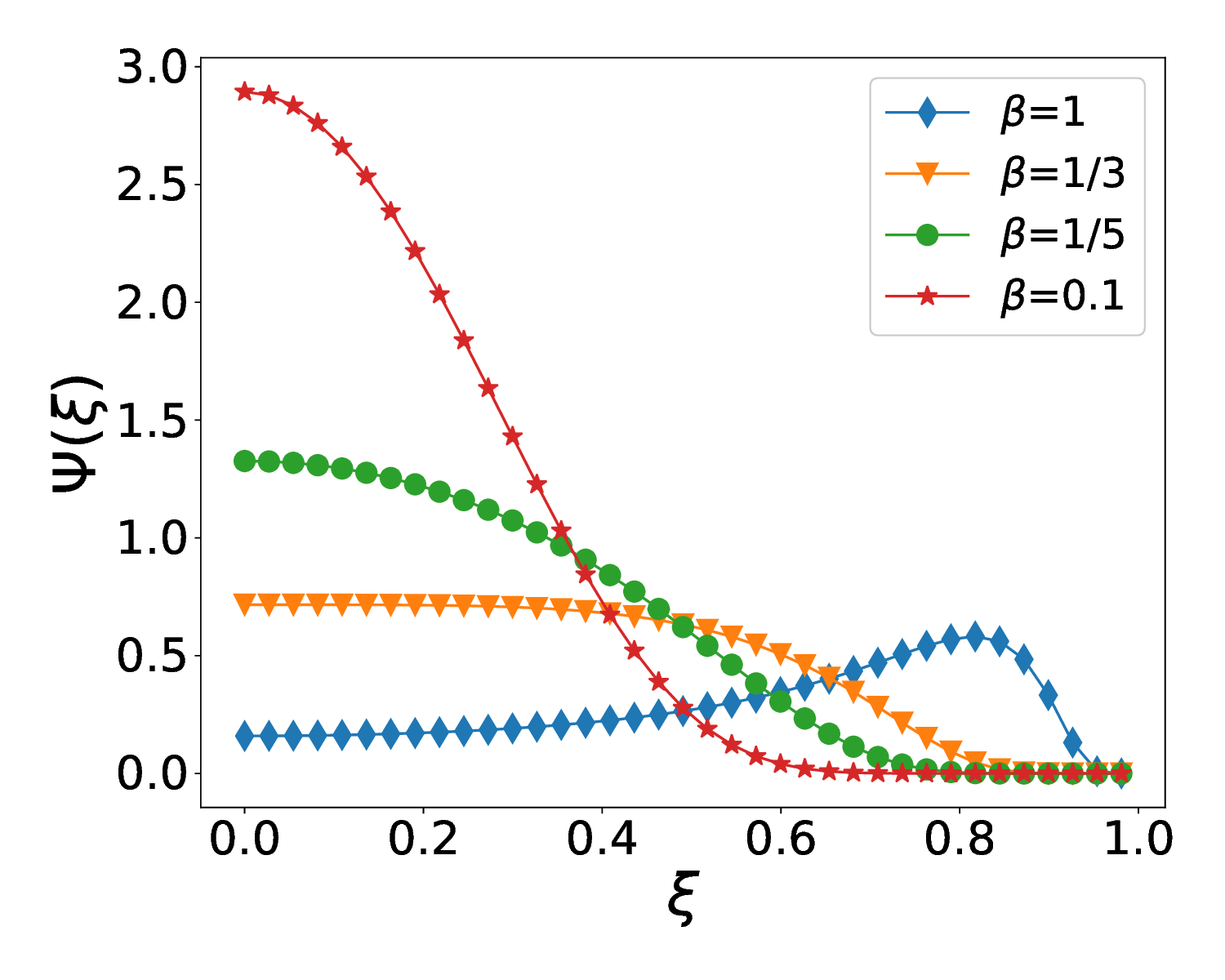}
\caption{The figure shows the distribution $\Psi(\xi)$ in Eq.~\ref{eq:Phi}, for various values of the ratio (Eq. \ref{eq:beta}). Observe the smooth transition from concave (red, stars) (peak at the center) to convex (blue, diamonds) (off-centered peak) shapes as $\beta$ increases.}
\label{fig:phi_theory}
\end{figure}
\begin{figure}
\centering
    \begin{minipage}{0.45\textwidth}
  \includegraphics[width=\linewidth]{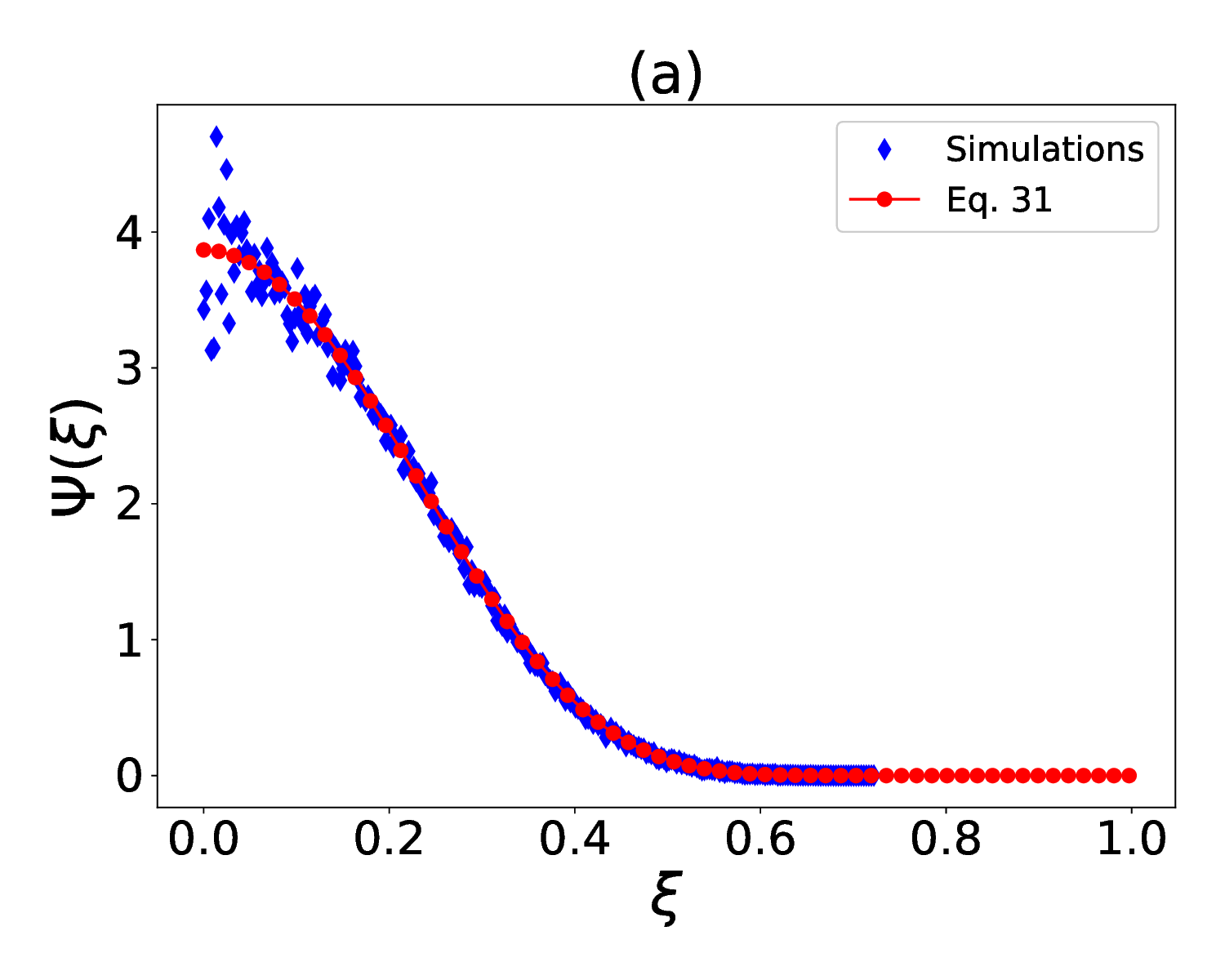}
\end{minipage} 
\begin{minipage}{0.45\textwidth}
  \includegraphics[width=\linewidth]{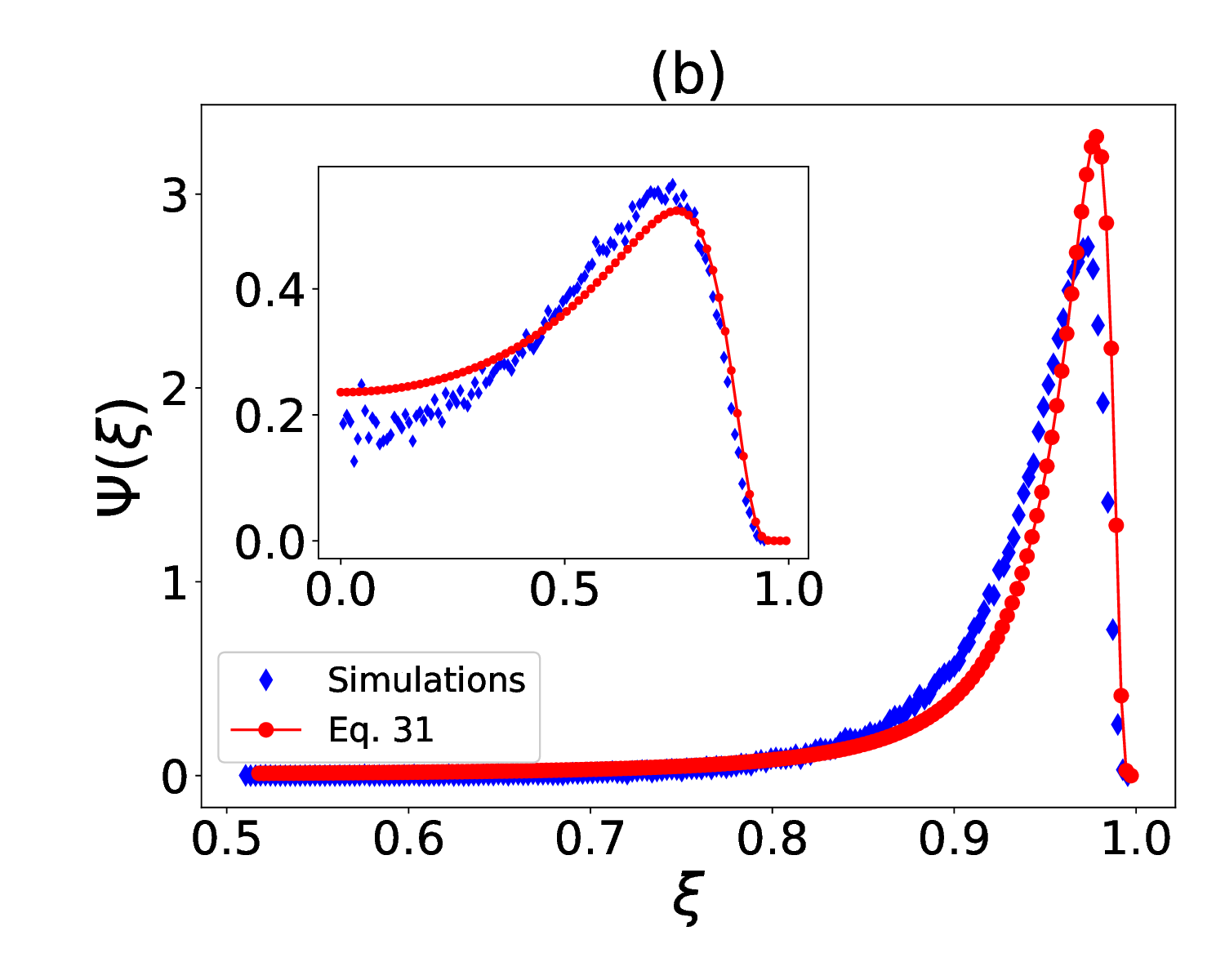}
\end{minipage}% <-- added
\caption{The analytical solution in Eq.~\ref{eq:Phi} is compared with the corresponding simulation results (obtained with $D_t=0, u_0>0\implies D_t^{\prime}=0$) for $\beta$ = 0.076 (a), $\beta$ =0.76 (inset of (b)) and $\beta$=7.64 (b), corresponding to three different values of $\kappa$ (see Table \ref{tab2}). The fixed parameters are given in Table~\ref{tab1}.}
\label{fig:phi_dt}
\end{figure}

% \begin{equation}
% \begin{aligned}
%     && u_0^2 \int \hat{\bf u} (\hat{\bf u}\cdot \nabla P) d^d \hat{\bf u}-k u_0 \int {\bf r}(\hat{\bf u} \cdot \nabla P) d^d \hat{\bf u} \\
%     && -k u_0\int \hat{\bf u} (\textbf{r}\cdot \nabla P)d^d\hat{\bf u}+ k^2 \int \textbf{r} (\textbf{r}\cdot \nabla P) d^d\hat{\bf u}\\
%     && -D_r \int [u_0 \hat{\bf u}-k \textbf{r}]\nabla_{\hat{\bf u}}^2 P]d^d\hat{\bf u}=0
% \end{aligned} 
% \label{eq:a1_main1}
% \end{equation}
\begin{equation}
 \begin{aligned}
       \int \hat{\bf u} (\hat{\bf u}\cdot \nabla P) d^d \hat{\bf u}- r_m^{-1}\bigg\{\int {\bf r}(\hat{\bf u} \cdot \nabla P) d^d \hat{\bf u}\\
      + \int \hat{\bf u} ({\bf r}\cdot \nabla P)d^d\hat{\bf u}\bigg\}+ r_m^{-2}\int {\bf r} (\textbf{r}\cdot \nabla P) d^d\hat{\bf u}\\
       -\frac{1}{\beta r_m}\int \bigg(\hat{\bf u}-\frac{\textbf{r}}{r_m}\bigg)\nabla_{\hat{\bf u}}^2 P d^d\hat{\bf u}=0
\end{aligned} 
\label{eq:a1_main1}
\end{equation}
where 
\begin{equation}
    \beta=\frac{k}{D_r}
    \label{eq:beta}
\end{equation}
is an important dimensionless parameter. The second term in the above equation can be simplified as follows: 
\begin{equation}
   \begin{aligned}
    \int {\bf r}(\hat{\mathbf{u}} \cdot \nabla P) d^d\hat{\mathbf{u}}= &\textbf{r} \nabla \cdot \{ \overline{\hat{\mathbf{u}}}(\mathbf{r}) \Phi(\mathbf{r}) \}\\
     =& \frac{\textbf{r}}{r_m} \left[(\nabla \cdot \textbf{r}) \Phi(\textbf{r})+(\textbf{r}\cdot \nabla \Phi)\right], \\
 \end{aligned} \label{eq:a1_term2}
\end{equation}
where we have made use of Eq.~\ref{eq:u_bar} for the average of the unit propulsion vector. The third term in Eq.~\ref{eq:a1_main1} is simplified as 
\begin{equation}
\begin{aligned}
     \int \hat{\bf u}(\textbf{r} \cdot \nabla P) d^d\hat{\bf u}= & (\textbf{r}\cdot \nabla)(\overline{\hat{\mathbf{u}}}(\mathbf{r})\Phi)\\
     =& \frac{\bf r}{r_m}[\Phi(\textbf{r})+\textbf{r}\cdot \nabla \Phi].
     \end{aligned}
     \label{eq:a1_term3}
\end{equation}
\begin{comment}
by making use of the vector identity 
\begin{equation*}
\begin{aligned}
    x_i \frac{\partial}{\partial x_i} (x_j \hat{e}_j \Phi)=& \hat{e}_j x_i \delta_{ij} \Phi +x_j \hat{e}_j x_i \frac{\partial}{\partial x_i} \Phi. 
    \end{aligned} 
\end{equation*}
\end{comment}
The fourth integral term in Eq.~\ref{eq:a1_main1} becomes 
$\textbf{r}(\textbf{r}\cdot \nabla \Phi)$ after the trivial angular integration. Finally, the fifth term in Eq.~\ref{eq:a1_main1} is simplified using the formula~\cite{celani2010bacterial} 
\begin{equation}
    \int {\bf \hat{u}} \nabla^2_{\bf \hat{u}}P d^d {\bf \hat{u}}=(1-d)\Phi({\bf r}) \overline{\hat{\mathbf{u}}}(\bf r), \label{eq:a1_term5}
\end{equation}
while $\int \nabla^2_{\bf \hat{u}}Pd^d{\bf\hat{u}}=0$ identically. Using equations Eq.~\ref{eq:a1_term2}, Eq.~\ref{eq:a1_term3} and Eq.~\ref{eq:a1_term5}, Eq.~\ref{eq:a1_main1} is simplified to 
\begin{equation}
    \int \hat{\bf u}(\hat{\bf u}\cdot \nabla P) d^d\hat{\bf u}+\frac{\bf r}{r_m^2}\bigg[\alpha\Phi-\textbf{r}\cdot \nabla \Phi]\bigg] =0 \label{eq:a1_main2}
\end{equation}
where 
\begin{equation}
   \alpha=\frac{(d-1)}{\beta}-(d+1).  
   \label{eq:alpha}
\end{equation}
is an important parameter in the problem. Mathematical calculations (in $d=2$) and numerical simulation results ($d=2$ and $d=3$) indicate that the positional distribution $\Psi(\xi)$ undergoes an interesting shape transition at $\alpha=0$, which is reminiscent of a continuous phase transition.
\subsubsection{Two dimensions: exact FPE and a Gaussian approximation}
So far, our treatment has been general, as far as the spatial dimensionality of the problem is concerned. For most of the remaining part of the paper, we will specialize to the dynamics of ABP in two dimensions, where there are only three degrees of freedom for the particle: the two plane polar coordinates ($r,\phi$) characterize the position vector ${\bf r}$, while the orientation angle $\theta_u$ characterizes the unit propulsion vector $\hat{\bf u}=(\cos\theta_u, \sin\theta_u)$. From Eq.~\ref{eq:langevin_main}, the dynamics of $\theta_u$ is given by the equation
\begin{equation}
        \frac{d \theta_u}{dt}=\sqrt{2D_r}\eta_{r}(t), 
   \label{eq:langevin1}
\end{equation}
where $\eta_{r}$ is uncorrelated Gaussian white noise with zero mean. 

 We will now carry out the simplification of the first term in Eq.~\ref{eq:a1_main2} in two dimensions, by using plane polar coordinates. On account of the radial symmetry, $P(r,\phi,\theta_u) \equiv P(r,\chi)$ depends only on the combination $\chi=\theta_u-\phi$~\cite{malakar2020steady}. To separate the variables, we introduce the conditional distribution $f(\chi|\xi)$ of the orientation $\chi$ at a given $r$: 
  \begin{equation}
   f(\chi|\xi)=\frac{P(r_m \xi,\chi)}{\Phi(r_m \xi)}.       \label{eq:r9}
  \end{equation} 
Note that $\hat{\bf u}=\cos{\chi}~\hat{\bf r}+\sin{\chi}~\hat{\boldsymbol \phi}$ and $\nabla P=\partial_r P~\hat{\bf r}+r^{-1}\partial_{\phi} P ~\hat{\boldsymbol \phi}$ in polar coordinates. Only the radial component of the first term in Eq.~\ref{eq:a1_main2} is non-zero, which turns out to be (see Appendix A, for details)
\begin{equation}
\begin{aligned}
    &\hat{\bf r}\cdot\int \hat{\bf u} (\hat{\bf u}\cdot \nabla P)~d^2\hat{\bf u}=  \bigg(\sigma^2_{\cos{\chi}} +\frac{r^2}{r_m^2}\bigg) \Phi^{\prime}(r)\\&+\Phi(r)\bigg( \partial_r \sigma^2_{\cos{\chi}} +\frac{4r}{r_m^2}+\frac{2\sigma^2_{\cos{\chi}}-1}{r}\bigg),\label{eq:first_term1}
    \end{aligned}
\end{equation}
where $\sigma^2_{\cos{\chi}}(r)$ is the variance of $\cos{\chi}$ at fixed radial distance $r$, calculated using the conditional probability density $f(\chi|r)$, while $d^2\hat{\bf u}\equiv d\theta_u$ and $\Phi^{\prime}(r)\equiv d\Phi/dr$. We have also used Eq.~\ref{eq:u_bar} to make the substitutions 
\begin{equation}
\langle \cos\chi\rangle=\frac{r}{r_m} ~~;~~\langle \sin\chi\rangle = 0. 
\label{eq:coschi}
\end{equation}
After using Eq.~\ref{eq:first_term1} in Eq.~\ref{eq:a1_main2}, we arrive at the following {\it exact} Fokker-Planck equation for the scaled radial coordinate, in the limit $D_t^{\prime}\to 0$: 
\begin{equation}
    \sigma^2_{\cos{\chi}} \frac{d\Psi}{d\xi}+\bigg[\partial_\xi \sigma^2_{\cos{\chi}} +\frac{2 \sigma^2_{\cos{\chi}}-1}{\xi}+\bigg(1+\frac{1}{\beta}\bigg)\xi\bigg]\Psi=0 \label{eq:diff_Phi2}
\end{equation}
The variance $\sigma^2_{\cos\chi}$ is still unknown in Eq.~\ref{eq:diff_Phi2}. While an exact result for the same is difficult to obtain, a Gaussian assumption for $\chi$ yields the following useful estimate, which we derive in Appendix B: 
 \begin{equation}
     \sigma^2_{\cos{\chi}}\simeq \frac{1}{2}( 1-\xi^2)^2, ~~~({\rm Gaussian}~ \chi) \label{eq:variance_cos}
 \end{equation}
After using Eq.~\ref{eq:variance_cos} in Eq.~\ref{eq:diff_Phi2}, we obtain the following closed form solution to Eq.~\ref{eq:diff_Phi2}: 
\begin{eqnarray}
\Psi(\xi)\simeq C (1-\xi^2)^{-3} \exp{\bigg(-\frac{1}{\beta(1-\xi^2)}\bigg)}~(\xi<1)~ \nonumber\\
\Psi(\xi)=0 ~~~~~~~~~~~~~~~~~~~~~~~~~~~~~~~~~~~~~~~~~(\xi\geq 1)~~
\label{eq:Phi}
\end{eqnarray}
 The normalisation constant is found to be $C= [\pi\beta(1+\beta)]^{-1}e^{\frac{1}{\beta}}$.

\noindent
{\it The ``active Gaussian" limit }: At small distances ($\xi\ll 1$), it is easily seen (by taking the logarithm of the l.h.s and subjecting the r.h.s to a binomial expansion) that the solution in Eq.~\ref{eq:Phi} reduces to the asymptotic Gaussian form
\begin{equation}
    \Psi(\xi)\sim \Psi(0) e^{-\alpha \xi^2},~~~~(\xi\ll 1)
    \label{eq:active_gaussian}
\end{equation}
which is a Boltzmann-like distribution, with effective active temperature 
\begin{equation}
    k_B T_a=\frac{\zeta_t u_0^2}{2k\alpha}, 
    \label{eq:Teff}
\end{equation}
with $\alpha$ given by Eq.~\ref{eq:alpha}. Following the fluctuation-dissipation relation, the corresponding ``active diffusion coefficient" is $D_a=u_0^2/2k\alpha$, which approaches the expected limiting form, $D_a\sim u_0^2/2D_r$ in the limit $\beta\ll 1$ ($k\ll D_r)$~\cite{basu2018active,basu2019long}. Further, in the limit $\kappa\to 0$ at fixed $D_r$, $r_m\to\infty$ (Eq.~\ref{eq:rm}), hence the range of validity of the asymptotic form  in Eq.~\ref{eq:active_gaussian} stretches to all $r$~\cite{chaudhuri2021active}. 
\begin{figure}
\centering
    \begin{minipage}{0.45\textwidth}
  \includegraphics[width=\linewidth]{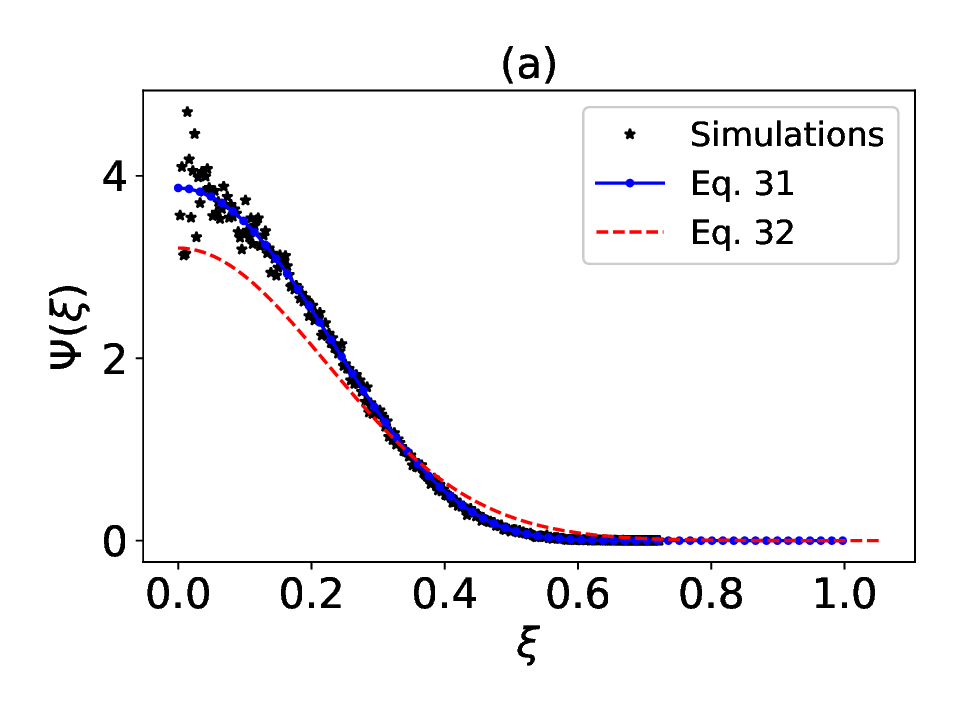}
\end{minipage} 
\begin{minipage}{0.45\textwidth}
  \includegraphics[width=\linewidth]{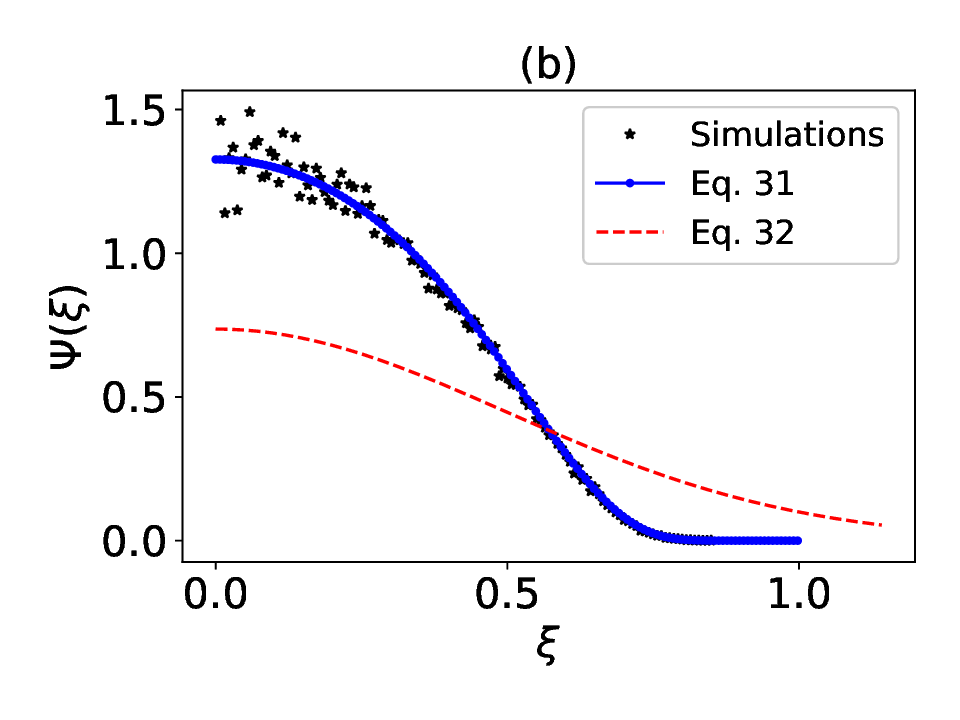}
\end{minipage}% <-- added
\caption{The analytical solutions in Eq.~\ref{eq:Phi} and Eq.~\ref{eq:active_gaussian} are compared with the corresponding simulation results (with $D_t=0, u_0>0$) for $\beta=0.076$ (a) and $\beta=0.2$ (b), corresponding to two different values of $\kappa$ (see Table \ref{tab2}). The fixed parameters are given in Table~\ref{tab1}. Note that, for the smaller $\beta$, the two different analytical expressions approach each other and are close to the simulation data. For the larger $\beta$, however, Eq.~\ref{eq:Phi} provides a much better fit than Eq.~\ref{eq:active_gaussian}.}
\label{fig:phi_dt0}
\end{figure}
\subsubsection{The concave-convex transition in $\Psi(\xi)$}
Fig.~\ref{fig:phi_theory} shows the plot of the function in Eq.~\ref{eq:Phi} as a function of the scaled variable $\xi$, for four different values of the dimensionless ratio $\beta$. The plot clearly shows a transition from concave (maximum at the origin) to convex (minimum at the origin) shapes as $\beta$ is increased. This behaviour of the distribution agrees with earlier experimental observations~\cite{takatori2016acoustic}. 
\begin{figure}[t]
\includegraphics[width=\linewidth]{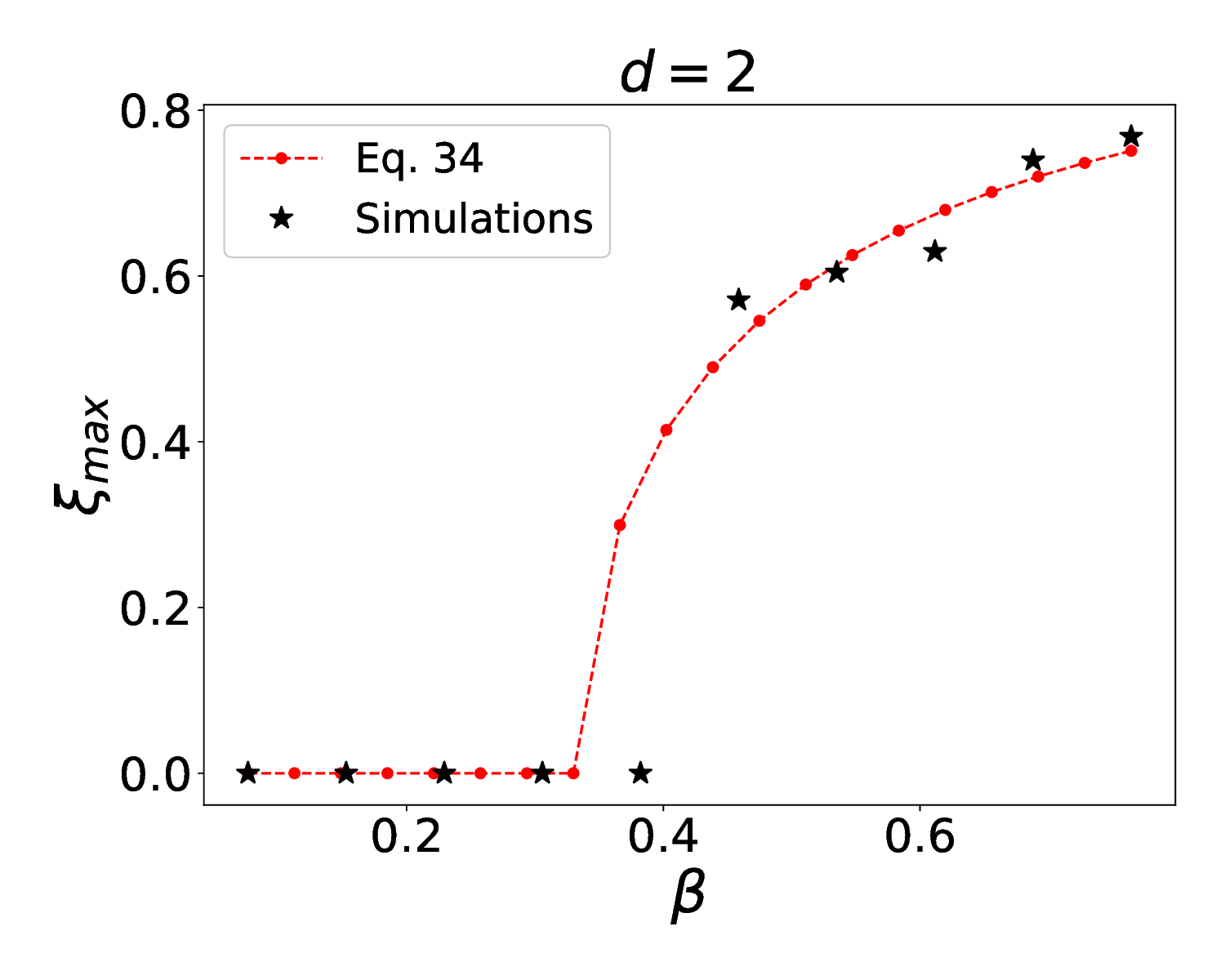}
\caption{For $d=2$, our theoretical prediction for the position $\xi_{\rm max}$ of the maximum of $\Psi(\xi)$, given by Eq.~\ref{eq:rmax} (red symbols) agrees well with simulation results (black symbols) obtained with $D_t=0, u_0>0\implies D_t^{\prime}=0$, with $\beta$ varied by changing $\kappa$ at fixed $D_r$ (see Table~\ref{tab2}). The critical value predicted theoretically is $\beta_c=1/3$. The fixed parameters are given in Table ~\ref{tab1}.} 
\label{fig:rmax_2d}
\end{figure} 
To understand better the nature of the transition between the concave and convex shaped distributions, we calculated the zeroes of the first derivative of $\Psi(\xi)$. It is easily found that $\xi=0$ and $\xi=\sqrt{1-1/3\beta}$ are the locations of the extrema of $\Psi(\xi)$, with the latter becoming imaginary (and hence irrelevant) for $\beta<\beta_c$ where $\beta_c=1/3$ is a ``critical value". It is also interesting to note that $\alpha=0$ for $\beta=\beta_c$, see Eq.~\ref{eq:alpha} (similar behaviour is observed in $d=3$ also, see the discussion following Eq.~\ref{eq:rma} in a later section). Hence, we conclude that the position of the peak of the distribution in Eq.~\ref{eq:Phi} is given by 
\begin{eqnarray}
    \xi_{\rm max}= 0~~~~~~~~~~~~\beta \leq \beta_c\nonumber\\
    \xi_{\rm max}=\sqrt{1-\frac{\beta_c}{\beta}}~~~\beta > \beta_c \label{eq:rmax}
\end{eqnarray}
In the vicinity of $\beta=\beta_c$, the second part of the above equation may be approximated by 

\begin{eqnarray}
    \xi_{\rm max}\sim \bigg(\frac{\beta-\beta_c}{\beta_c}\bigg)^{\frac{1}{2}}~~~\beta \gtrsim \beta_c \label{eq:rmax1}
\end{eqnarray}
Eq.~\ref{eq:rmax1} is reminiscent of critical behaviour in a continuous phase transition and predicts that the probability distribution $\Psi(\xi)$ undergoes a continuous change between concave and convex shapes (around the origin) as the ratio $\beta$ is increased beyond the critical value $\beta_c=1/3$. Here, it is important to note that both the concave and convex-shaped distributions are non-Gaussian in the limit $D_t^{\prime}\to 0$. As $D_t^{\prime}$ is increased, we expect both these distributions to 
cross over to the common, asymptotic Gaussian form (see Eq.~\ref{eq:formalsolution}). This transition is studied in detail in the next section, but some general arguments are presented in the following subsection. 

\subsection{Non-zero translational diffusion: $D_t^{\prime}>0$}
We now attempt to develop a qualitative understanding of the cross-over to equilibrium Gaussian (Boltzmann) distribution as $D_t^{\prime}$ is increased. 
\begin{figure*}[t]
\centering
    \begin{minipage}{0.45\textwidth}
  \includegraphics[width=\linewidth]{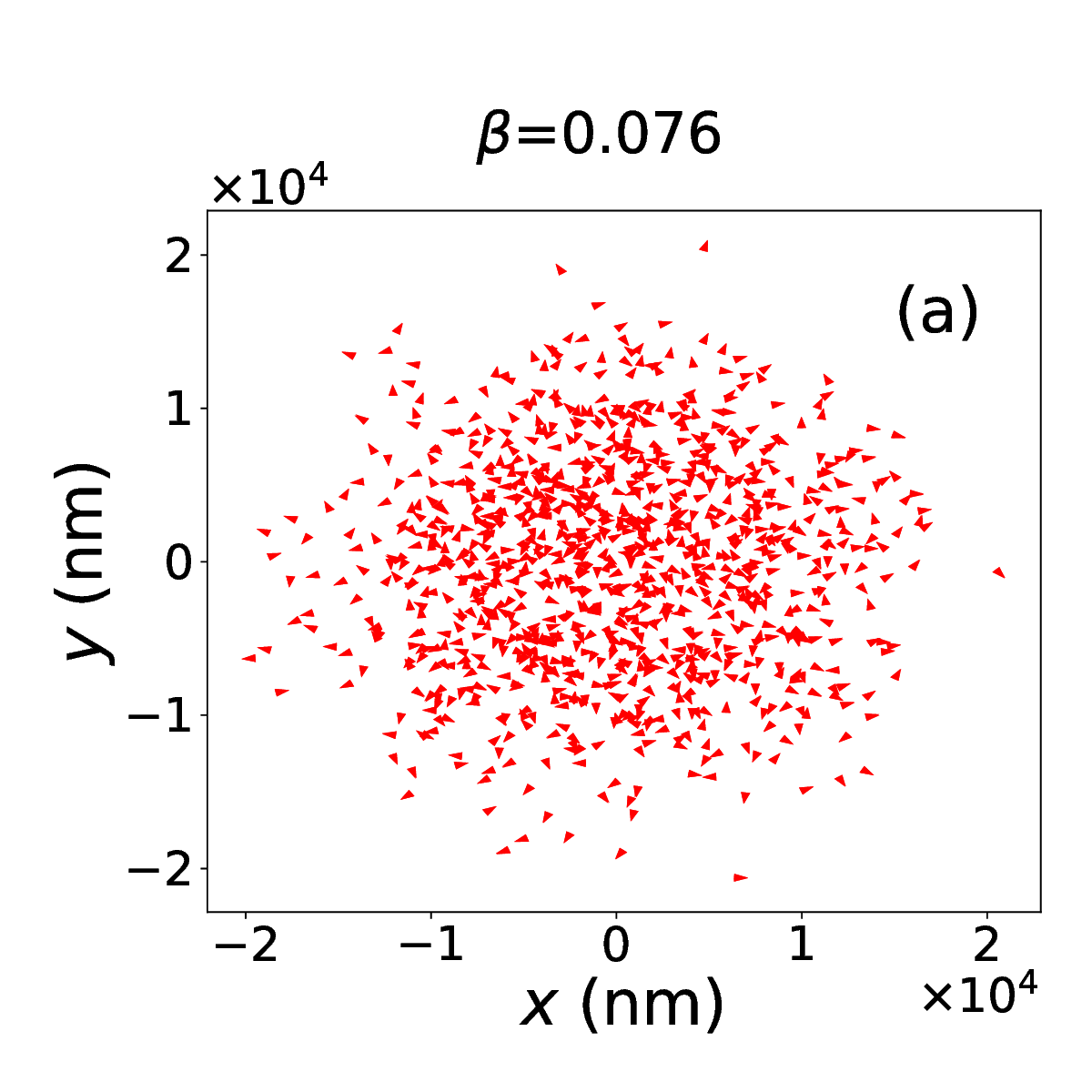}
\end{minipage} 
\begin{minipage}{0.45\textwidth}
  \includegraphics[width=\linewidth]{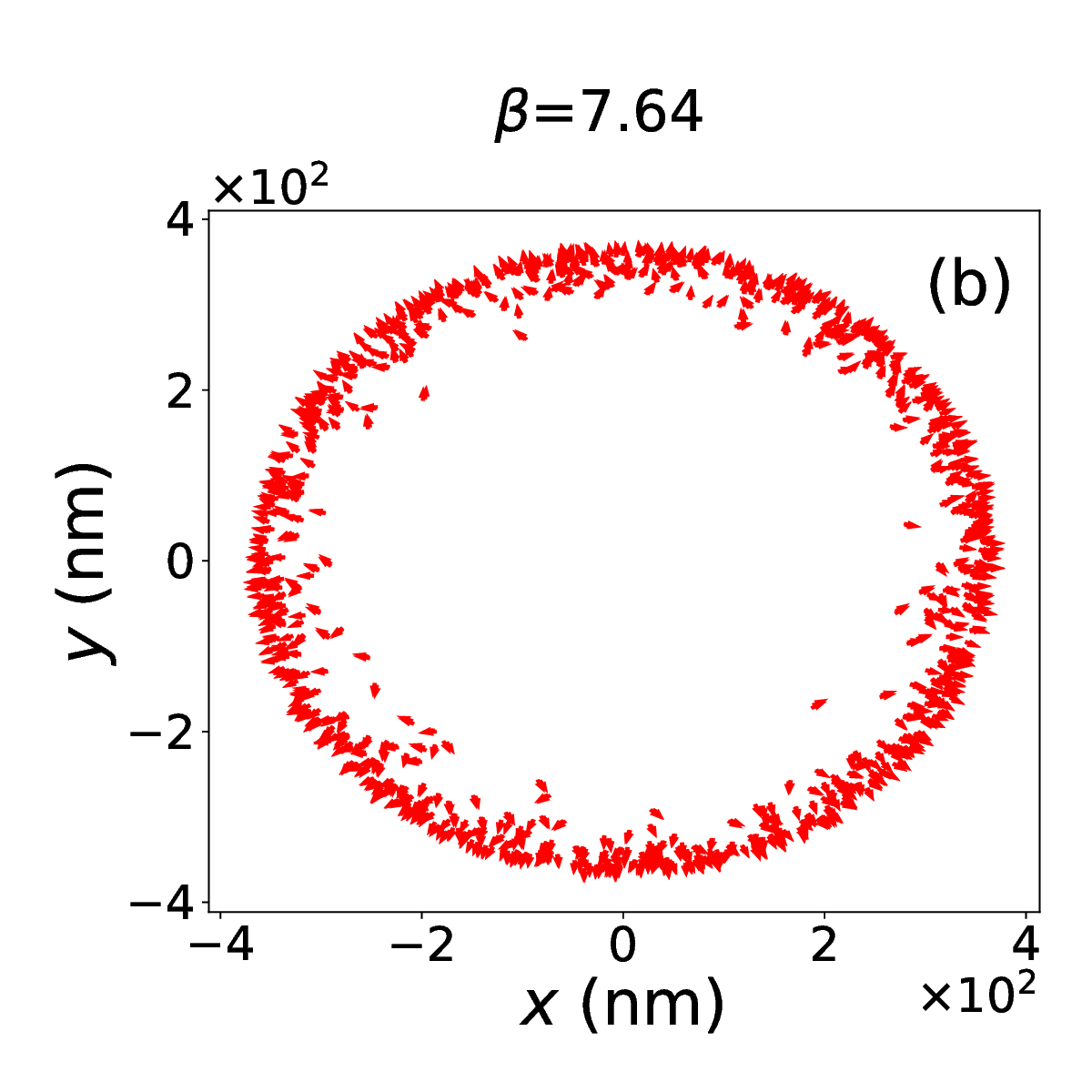}
\end{minipage}% <-- added
\caption{Simulation results for orientational profile on x-y plane of active velocity $\hat{\mathbf{u}}$ of ABP in $d=2$ for two values of $\beta$, realized by changing $\kappa$ at fixed $D_r$. See Table ~\ref{tab1} and Table ~\ref{tab2}. }
\label{fig:vector_plot_2d}
\end{figure*}

For large $D_t^{\prime}$, it is obvious from Eq.~\ref{eq:formalsolution} that $\Phi(r)$ approaches the equilibrium Boltzmann distribution, with peak at $r=0$.  The nature of the function $\Phi(r)$ in the vicinity of $r=0$ is directly related to the variance $\sigma^2_{\cos\chi}(0)$. To see this, consider Eq.~\ref{eq:dimensionless}, from which the first derivative of $\Psi(\xi)$ is always seen to vanish at $\xi=0$,  since $g(0)=0$ by symmetry. The second derivative at $r=0$ is found to be 
\begin{equation}
    \Psi^{\prime\prime}(0)=\frac{\Psi(0)}{D_t^{\prime}}(g^{\prime}(0)-1). 
    \label{eq:second-derivative}
\end{equation}
Therefore, the sign of the second derivative depends on whether $g^{\prime}(0)$ is less than or greater than 1. We prove the following result in Appendix C (Eq. \ref{eq:fprime1}: 
\begin{equation}
    g^{\prime}(0)=\frac{2\sigma^2_{\cos\chi}(0)-1}{D_t^{\prime}}.
    \label{eq:fprime}
\end{equation}
In Appendix D, we show that $\sigma^2_{\cos\chi}(0)=1/2$ for $D_t^{\prime}>0$ as well, therefore, Eq.~\ref{eq:fprime} leads to the condition $\Psi^{\prime\prime}(0) < 0$ (Eq.~\ref{eq:second-derivative}), for $D_t^{\prime}>0$. The implication is that, for $D_t^{\prime}>0$, origin is always a local maximum  for any value of $\beta$, and may be expected to become the dominant maximum with increase in $D_t^{\prime}$, for fixed $\beta$ (the ``active to passive transition"). Our simulations (see next section) give some supportive evidence for this picture, Indeed, by matching the second derivative at the origin, we also find from Eq.~\ref{eq:second-derivative} that $\Psi(\xi)\sim \Psi_{\rm eq}(\xi)$ near $\xi=0$, where $\Psi_{\rm eq}(\xi)\propto \exp(-\xi^2/2D_t^{\prime})$ is the equilibrium distribution in dimensionless form (Eq.~\ref{eq:psi}). We conclude that both the convex and concave distributions asymptotically tends to the equilibrium Gaussian form smoothly with increase in $D_t^{\prime}$.

We will now present results from numerical simulations, with detailed comparisons with the mathematical predictions wherever applicable. 
\section{\label{sec:level3}Numerical Simulation Results}
\begin{comment}
Brownian dynamics simulations of the problem were carried out in a two-dimensional plane, where the potential energy is  $U(\mathbf{r})=(\kappa/2)r^2$ where $r^2=x^2+y^2$. The Langevin equations in Eq.~\ref{eq:langevin_main} were simulated numerically using a first order integration methods (forward Euler discretization scheme), with averaged over $10^{5}$ realizations. Differential equations for variaables ($x,y,\theta$) solved by evalualting at discrete time steps, say $(x_i,y_i,\theta_i) \sim (x[t_i],y[t_i],\theta[t_i])$ and time step is given as $\Delta_i=i~\Delta t$, $\Delta t=10^{-3}$~s is sufficiently small. where The numerical results are obtained by solving the following equations,

\begin{equation*}
    \begin{aligned}
       & x_{i+1}=x_i+(u_0 \cos{\theta_i}-k x_i) \Delta t +\sqrt{2D_t \Delta t} ~\eta_{x,i} \\
        & y_{i+1}=y_i+(u_0 \sin{\theta_i}-k x_i) \Delta t +\sqrt{2D_t \Delta t} ~\eta_{y,i} \\
        & \theta_{i+1}=\theta_i+\sqrt{2 D_r \Delta t}~ \eta_{\theta,i}
    \end{aligned}
\end{equation*}
\end{comment}
Brownian dynamics simulations were performed in a two-dimensional plane using Langevin equations and a first-order integration method, with $10^{5}$ realizations averaged. The potential energy $U(\mathbf{r})=(\kappa/2)r^2$, where $r^2=x^2+y^2$, and the differential equations for the variables ($x$, $y$, $\theta_u$) were solved using the following discretization scheme, with $\Delta t=10^{-3}$ s: 
\begin{equation}
    \begin{aligned}
&x_{i+1}=x_i+(u_0 \cos{\theta_i}-k x_i) \Delta t +\sqrt{2D_t \Delta t} ~\eta_{tx,i},\\ &y_{i+1}=y_i+(u_0 \sin{\theta_i}-k y_i) \Delta t +\sqrt{2D_t \Delta t} \eta_{ty,i},\\ &\theta_{u,i+1}=\theta_{u,i}+\sqrt{2 D_r \Delta t} \eta_{rx,i},
 \end{aligned}
\end{equation}
where $\{x_i,y_i,\theta_{u,i}\} \equiv \{x(t_i),y(t_i),\theta_u(t_i)\}$ and $\eta_{tx,i}$, $\eta_{ty,i}$ and $\eta_{rx,i}$ are uncorrelated random numbers generated from independent Gaussian distributions, with zero mean and unit standard deviation. Here, subscripts $t$ and $r$ represent translational and rotational motion, respectively. 

 The active velocity $u_0$ was varied in the range $100$ $\mu$m s$^{-1}-$24 mm s$^{-1}$ \cite{bechinger2016active}. However, for most of the simulations (unless specifically mentioned), we used the value $u_0=3.67$ $\mu$m s$^{-1}$ (fixed arbitrarily). For characterising viscous damping in translational motion, we fixed $\zeta_t=10^{-5}$ pNs nm$^{-1}$, appropriate for a spherical particle with radius $a\simeq 0.53$ $\mu$m (similar to the size of a typical colloidal particle) in water. At $T=300$K, the corresponding translational and rotational diffusivities of the ABP turn out to be $D_t=0.4$ $\mu$m$^{2}$ s$^{-1}$ and $D_r=1.308$ s$^{-1}$ respectively. For trap stiffness $\kappa$, we chose values in the range $10^{-6}-10^{-4}$ pN nm$^{-1}$ such that both the regimes $\beta<\beta_c$ and $\beta>\beta_c$ are accessed in simulations. These ``standard" values are listed in Table~\ref{tab1}. 

In order to study the behaviour of the system in the limit $D_t^{\prime}\to 0$, we also did a few simulations with $D_t=0$, and $\kappa,u_0>0$ ($u_0$ fixed at the standard value in Table~\ref{tab1}, and $\kappa$ varied in the range mentioned above). From the simulations, we obtained the positional probability distribution function $\Psi(\xi)=r_m^2 \Phi(r_m \xi)$ and the conditional angular probability distribution $f(\chi|\xi)$ (Eq.~\ref{eq:r9}). The peak $\xi_{\rm max}$ of the positional distribution was obtained from $\Psi(\xi)$. The first and second moments of $\cos\chi$ were obtained as functions of $\xi$ from the conditional angular distribution $f(\chi|\xi)$. 
%\noindent

\begin{figure*}[t]
    \centering % <-- added
    \begin{minipage}{0.33\textwidth}
  \includegraphics[width=\linewidth]{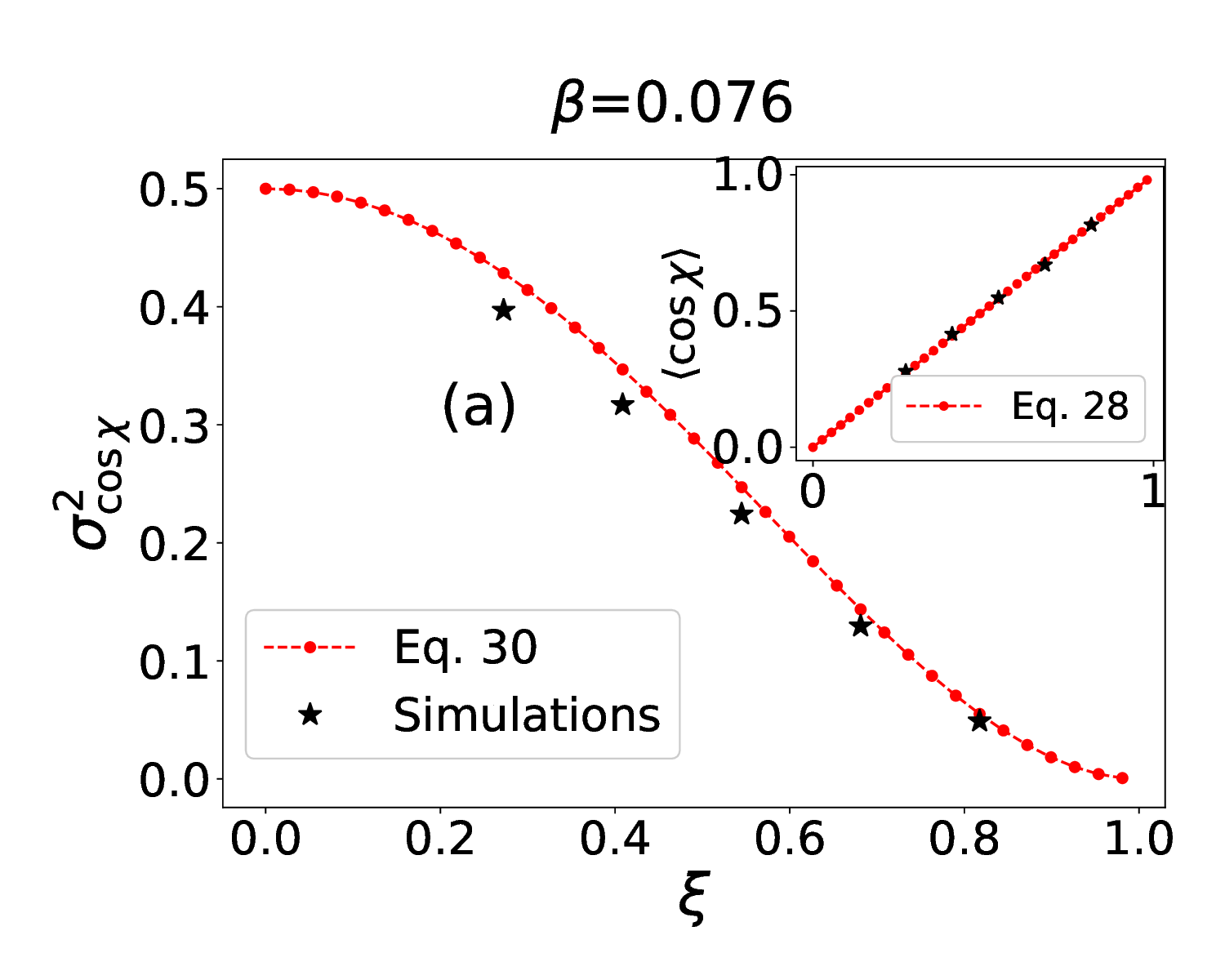}
  %\label{fig:plot_var2dt0}
\end{minipage} %\hspace{0.01pt}% <-- added
\begin{minipage}{0.33\textwidth}
  \includegraphics[width=\linewidth]{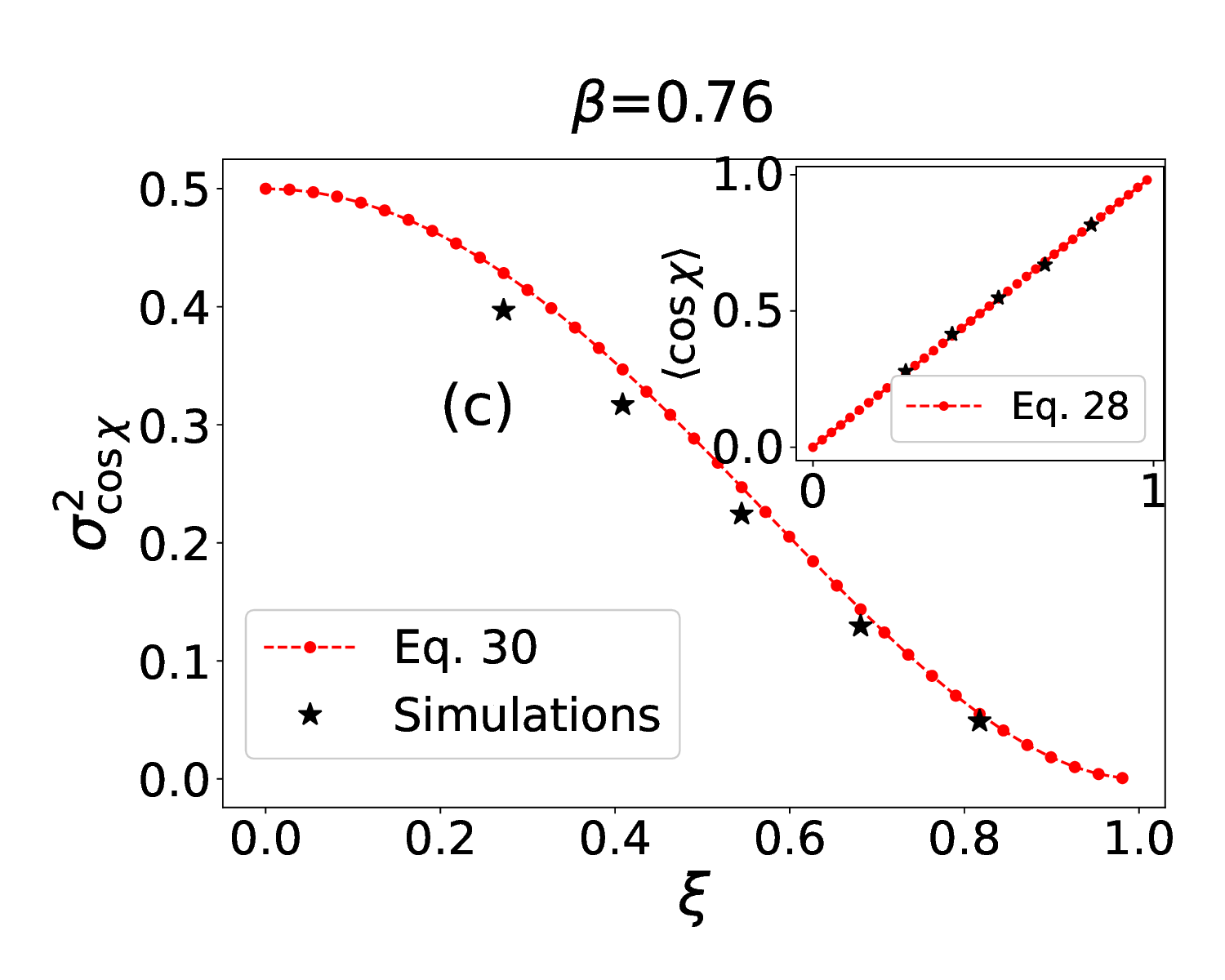}
  %\label{fig:plot_var1dt0}
\end{minipage}%\hspace{0.01pt}% <-- added
\begin{minipage}{0.33\textwidth}
  \includegraphics[width=\linewidth]{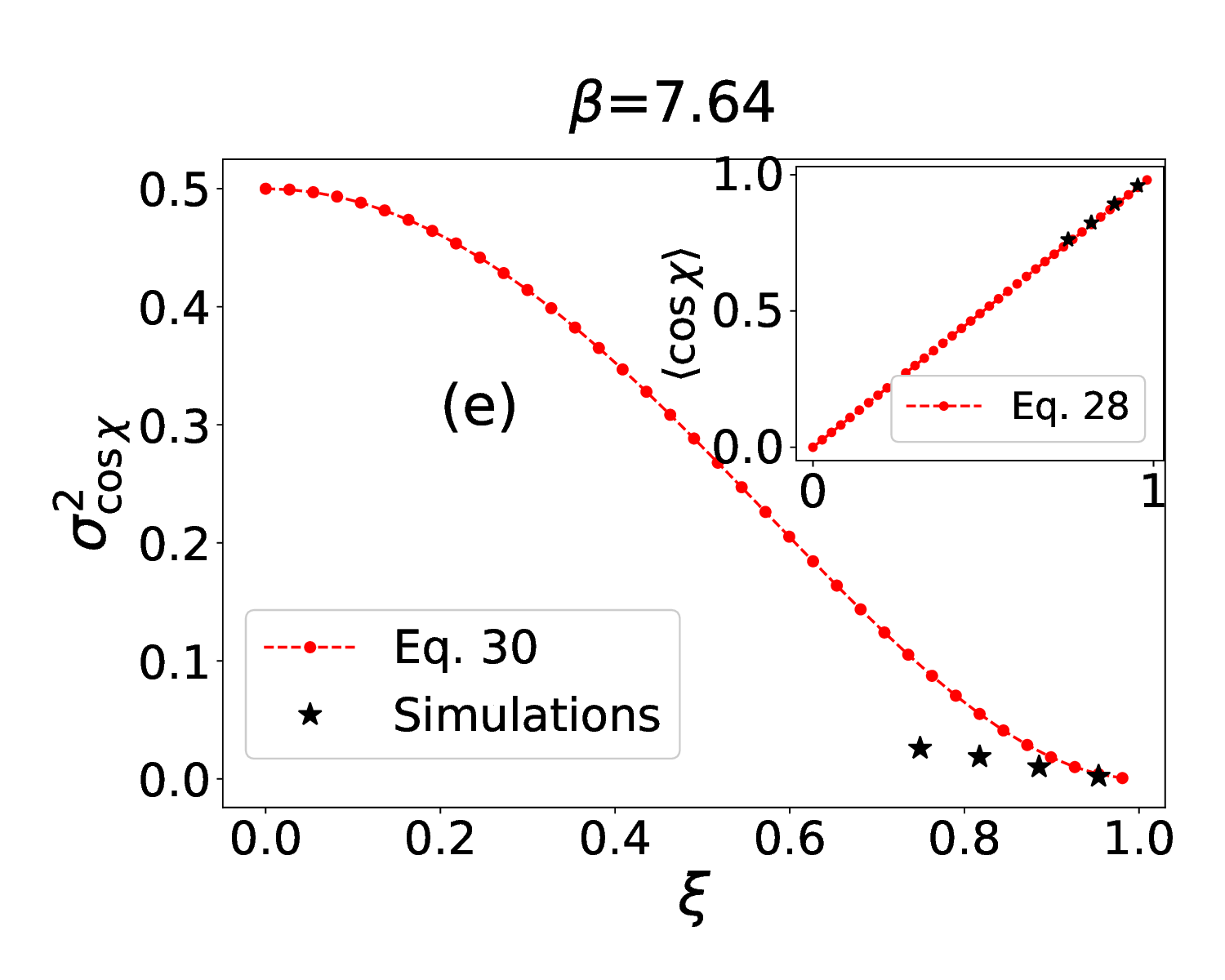}
  %\label{fig:plot_var0dt0}
\end{minipage} % <-- added
\vspace{0.1pt}
\begin{minipage}{0.33\textwidth}
  \includegraphics[width=\linewidth]{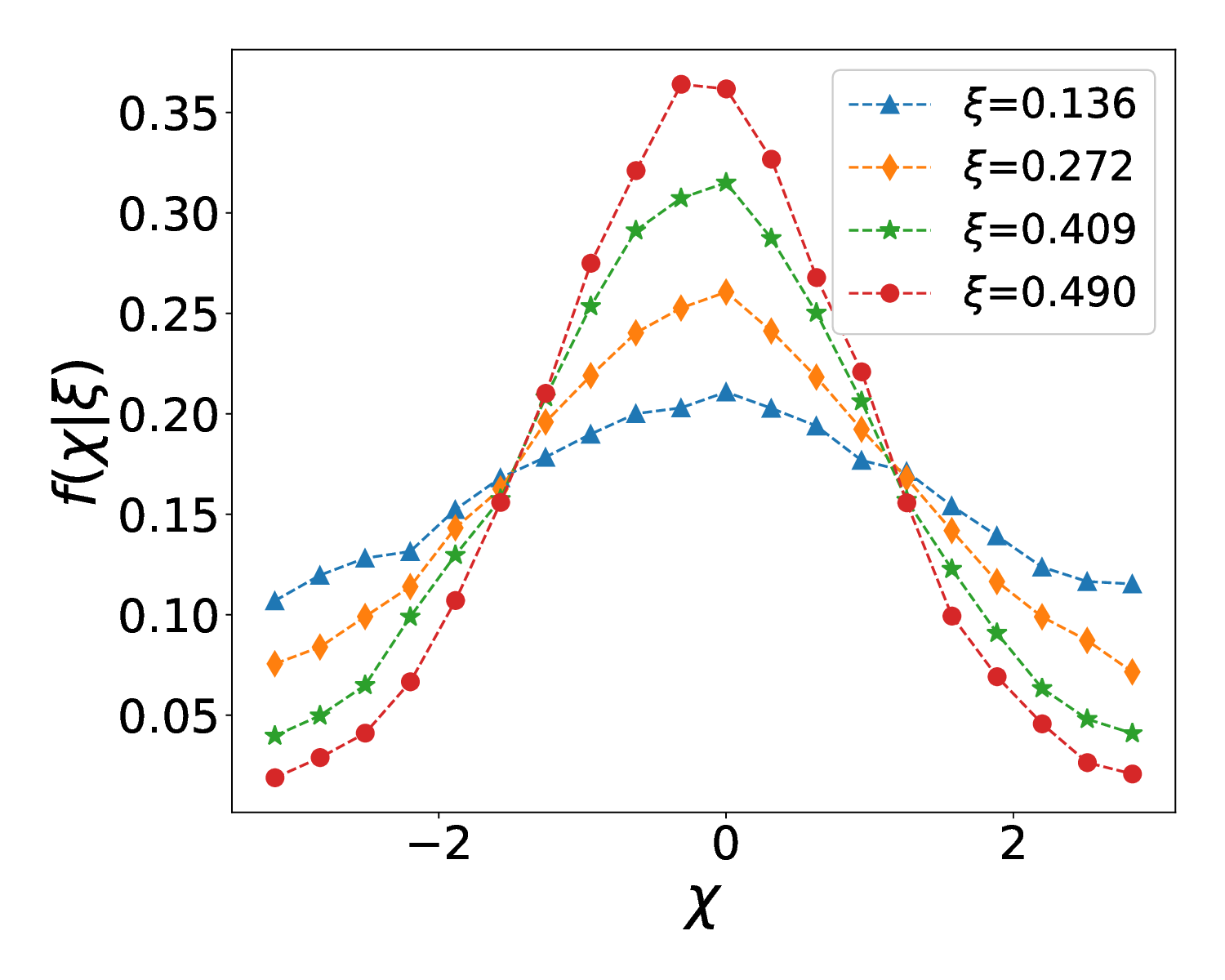}
  %\label{fig:plot_chi2dt0}
\end{minipage} % <-- added
\begin{minipage}{0.33\textwidth}
  \includegraphics[width=\linewidth]{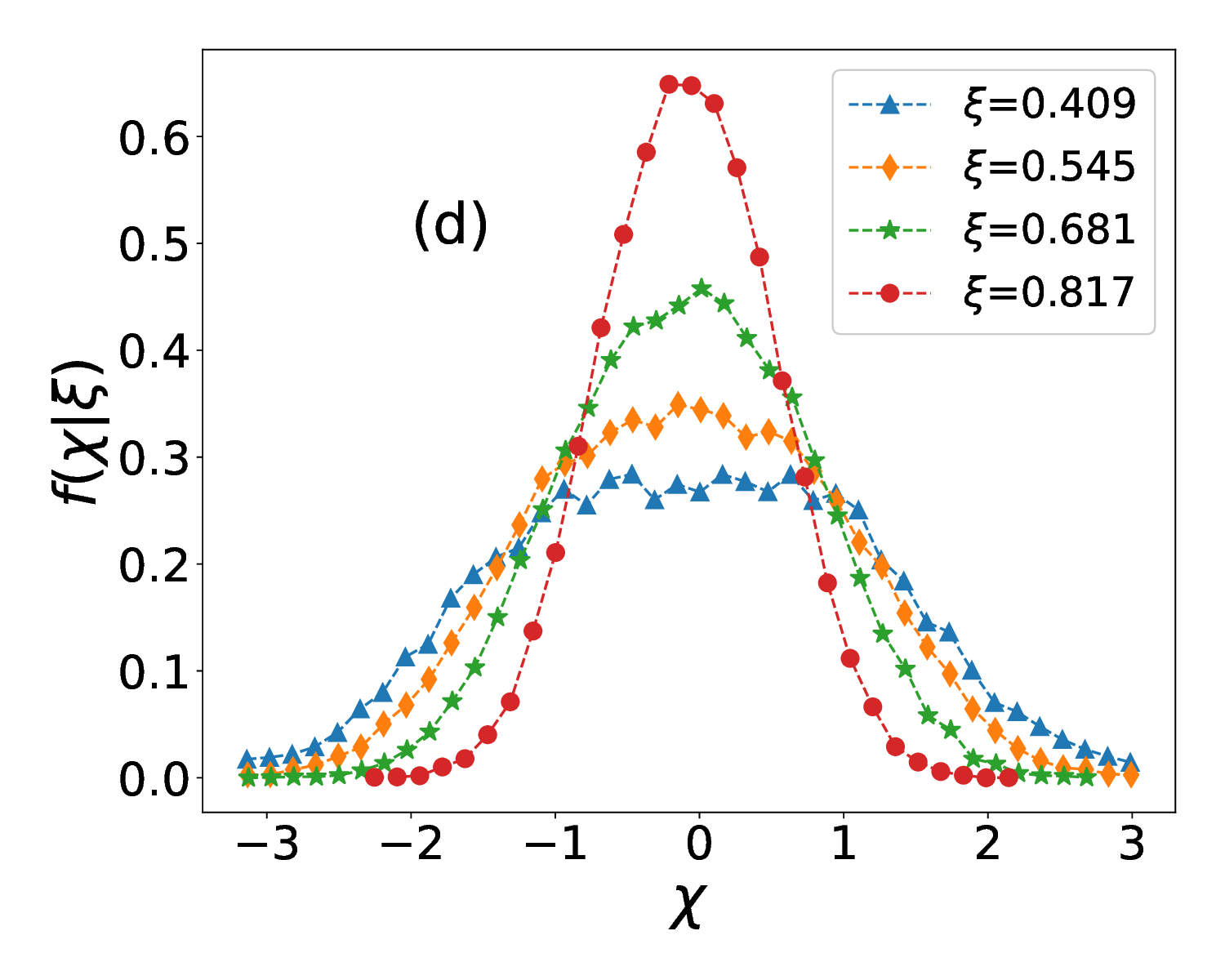}
 % \label{fig:plot_chi1dt0}
\end{minipage}% <-- added
\begin{minipage}{0.33\textwidth}
  \includegraphics[width=\linewidth]{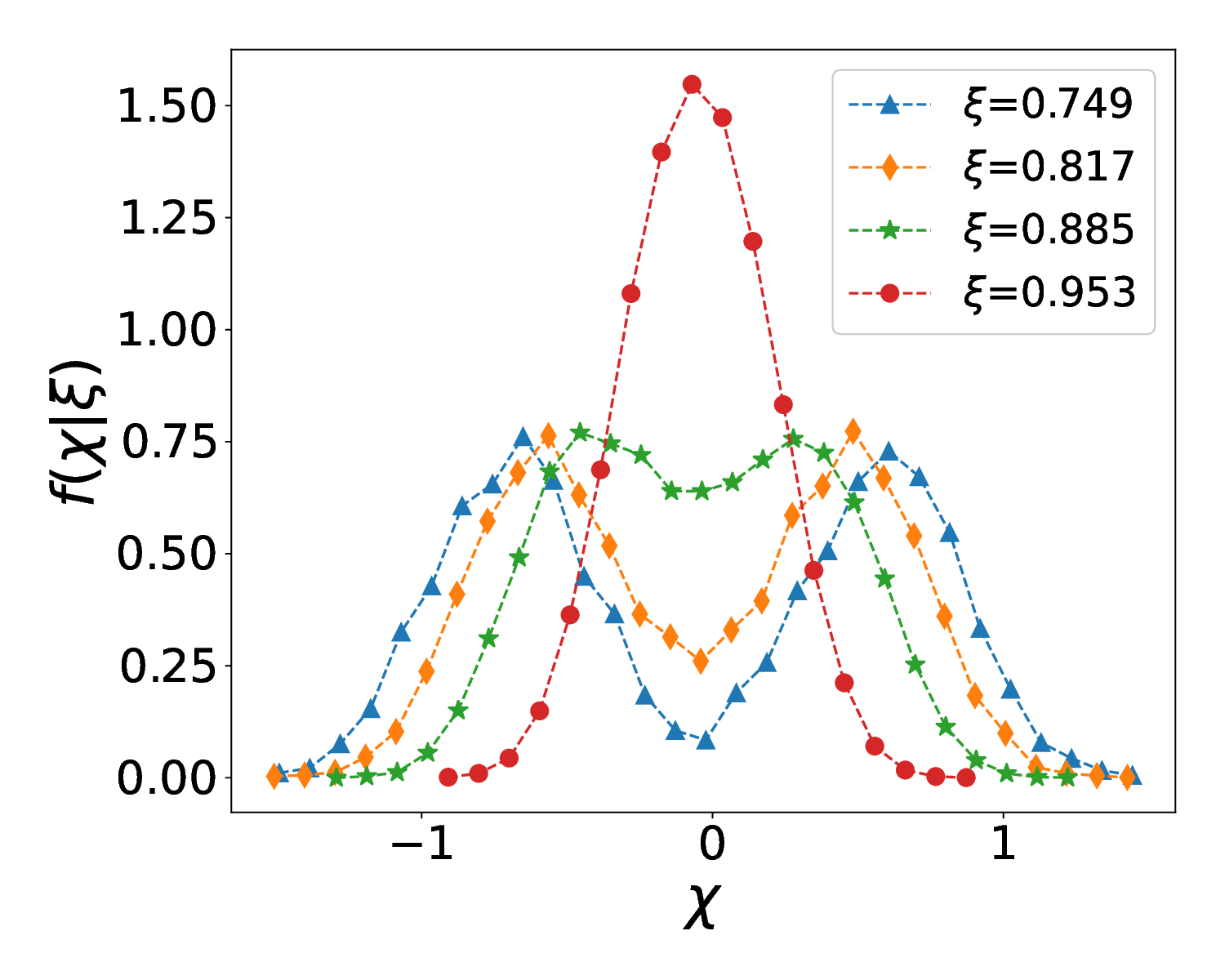}
  %\label{fig:plot_chi0dt0}
\end{minipage} 
\caption{The first row shows the results for mean(inset) and variance of ${\rm cos}\chi$, given along with the theoretical predictions from Eq.~\ref{eq:coschi} and Eq.~\ref{eq:variance_cos}, respectively. The figures in the second row show the distribution of orientation angle $\chi$ of the ABP at various radial distances from the trap centre (expressed as the scaled variable $\xi$), when $D_t^{\prime}=0~(D_t=0, u_0>0$). These figures correspond to different values of the $\beta$: (a,b) $\beta=0.076$, (c,d) $\beta=0.76$, and (e,f) $\beta=7.64$, the corresponding $\kappa$ values are given in Table \ref{tab2}. Note that the distribution changes its nature from unimodal to bimodal, with increase in $\beta$. The fixed parameters are given in Table~\ref{tab1}.}
\label{fig:coschi_dt0}
\end{figure*}

\subsection{$D_t^{\prime}=0$, The concave-convex transition in $\Psi(\xi)$}
For the  special case $D_t^{\prime}=0$ (to simulate this limit, we put $D_t=0$ and kept $u_0>0$, see Table ~\ref{tab1}), our simulation results for the probability distribution $\Psi(\xi)$ show excellent agreement with the mathematical expression derived in Eq.~\ref{eq:Phi}, as is observed from Fig.~\ref{fig:phi_dt}. Notably, the distribution shows a transition in shape, from concave (maximum at the origin) to convex (minimum at the origin, maximum away from the origin) forms, as the trap stiffness is increased. This non-intuitive, strongly non-equilibrium behaviour has been reported in several experimental~\cite{takatori2016acoustic,schmidt2021non,dauchot2019dynamics} and theoretical papers~\cite{pototsky2012active,basu2018active,solon2015active}. In their experiments with Janus particles under optical tweezers, Takatori et al.~\cite{takatori2016acoustic} observed a transition in the particle density profile, from a state with maximum particle density at the trap center($\xi=0$) to a distribution peaked at $\xi \sim 1$, as the trap stiffness increases. Similarly, in another, more recent experiment~\cite{schmidt2021non}, at low laser powers, the distribution of nano-particles was found to be characterised by a Gaussian density profile and as the laser power increases the particle's density acquires a non-Gaussian form. A few theoretical papers~\cite{pototsky2012active,malakar2020steady, basu2019long,chaudhuri2021active} have studied the crossover between  Gaussian and non-Gaussian positional distributions in ABP. Stark et al.~\cite{pototsky2012active} studied the stationary state of active particles in the limit of small and large rotational diffusivity and observed similar crossover in the positional distribution of ABP. Later on, they used this model to understand the collective behaviour of active particles.
Malakar et al.~\cite{malakar2020steady} obtained an exact perturbative solution for $\Psi(\xi)$ as a power-series expansion in a dimensionless P\'{e}clet number $u_0/\sqrt{D_tD_r}$, from which useful limiting forms were obtained in various cases, including the limit $D_t^{\prime}\to 0$. Basu et al.~\cite{basu2019long} derived exact limiting forms of $\Psi(\xi)$  in the limits $D_r\to 0$ and $D_r\to \infty$. In comparison, our result in Eq.~\ref{eq:Phi} provides an explicit form of the positional distribution in the limit $D_t^{\prime}\to 0$, predicts the existence of distinct ``concave'' and ``convex'' phases and the nature of the transition between them. 

In Fig.~\ref{fig:phi_dt0}, the concave shaped distributions for two sub-critical values of $\beta$ are plotted as function of the scaled radial distance $\xi$. The fits show that the theoretical prediction in Eq.~\ref{eq:Phi} fits the data very well in both cases, while the limiting ``active" Gaussian function in Eq.~\ref{eq:active_gaussian} fits the same reasonably well for the smaller value of $\beta$, in agreement with our expectations. 

In Fig.~\ref{fig:rmax_2d}, we plot the maximum $\xi_{\rm max}$ (as the dimensionless ratio) of the distribution (plotted in Fig.~\ref{fig:phi_dt}), against $\beta$. The corresponding theoretical prediction in Eq.~\ref{eq:rmax} is also shown for comparison, and both show excellent agreement with each other. 

\begin{figure*}[t]
    \centering % <-- added
    \begin{minipage}{0.33\textwidth}
  \includegraphics[width=\linewidth]{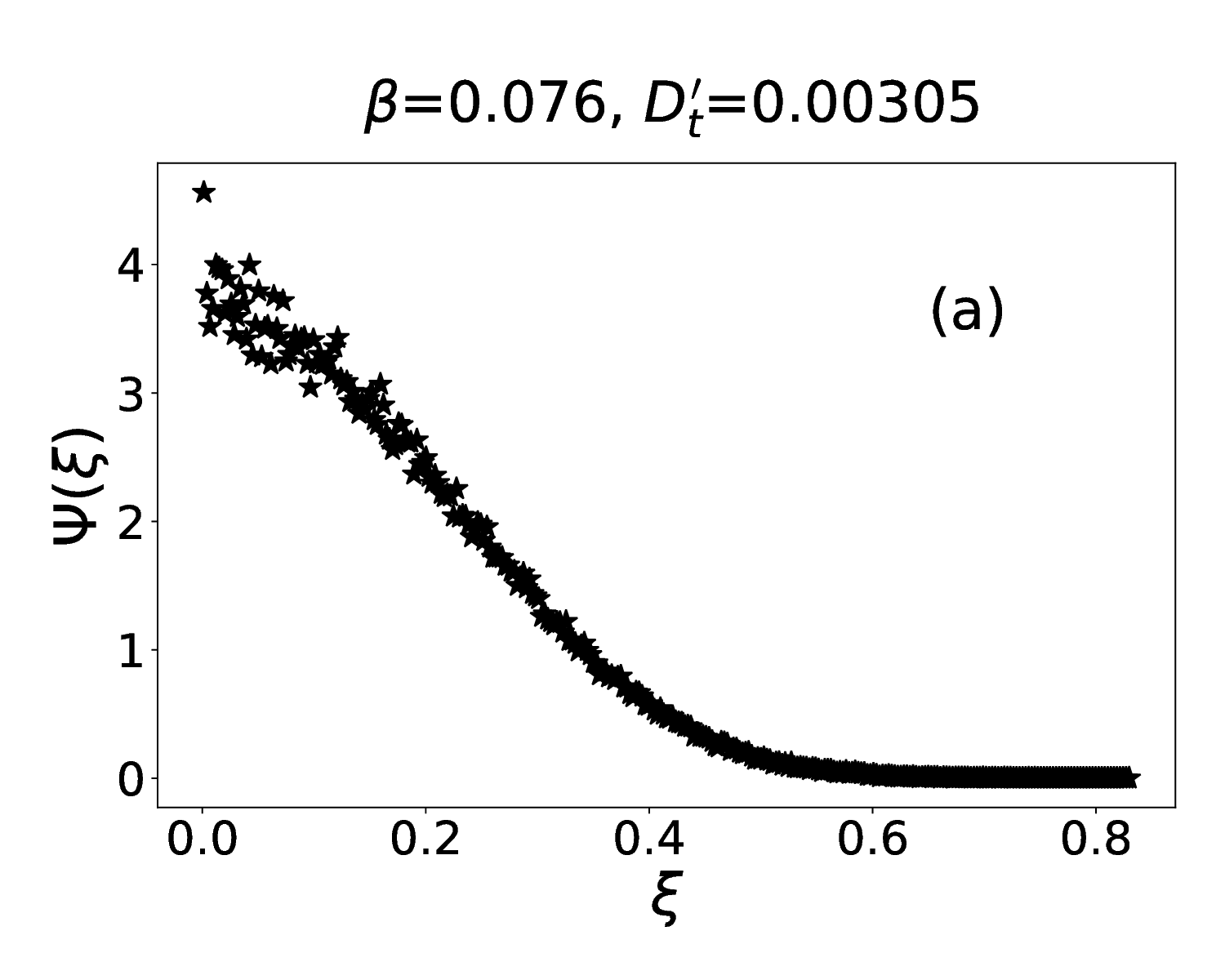}
  %\label{fig:plot_var2dt0}
\end{minipage} \hspace{0.01pt}% <-- added
\begin{minipage}{0.33\textwidth}
  \includegraphics[width=\linewidth]{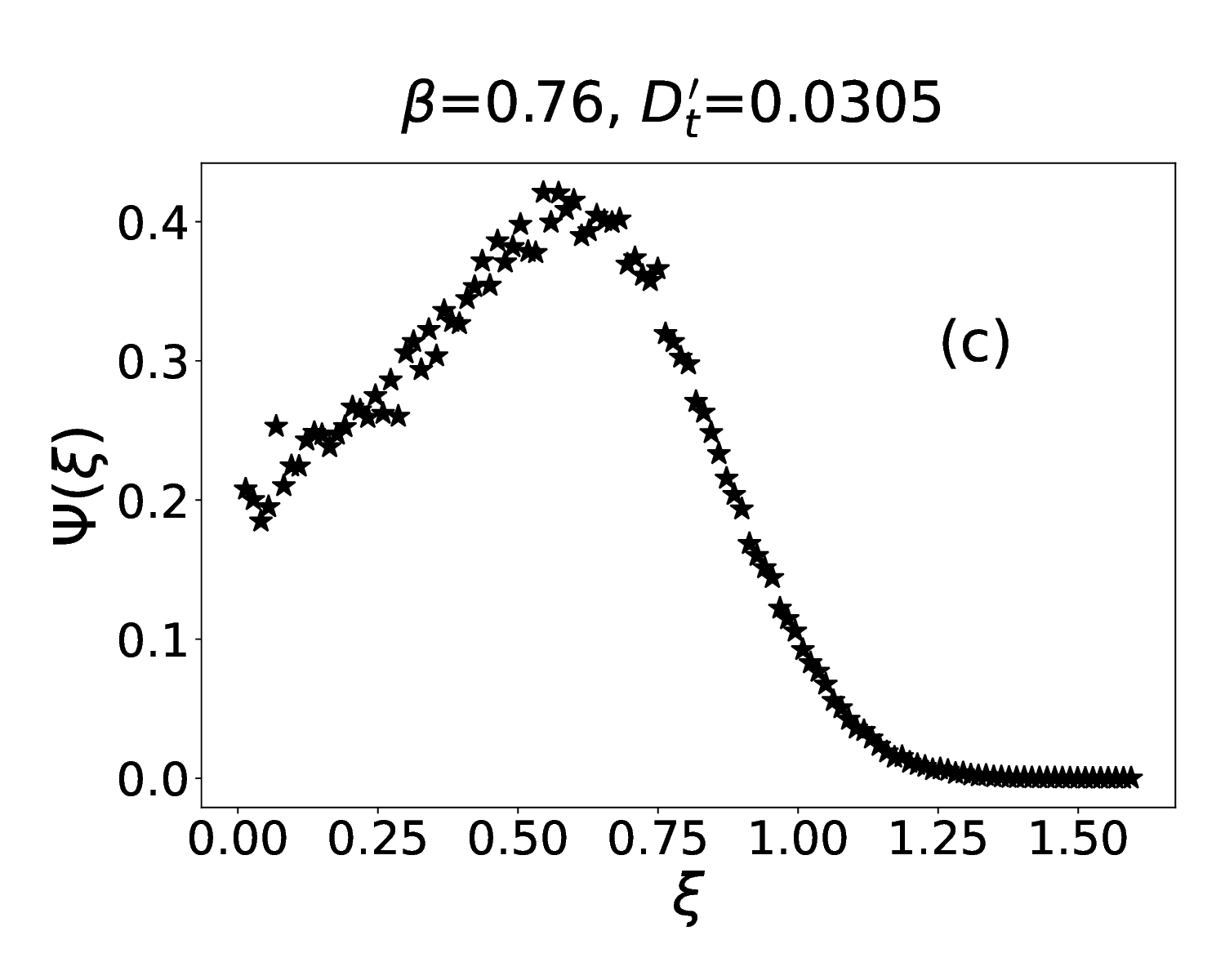}
  %\label{fig:plot_var1dt0}
\end{minipage}\hspace{0.01pt}% <-- added
\begin{minipage}{0.33\textwidth}
  \includegraphics[width=\linewidth]{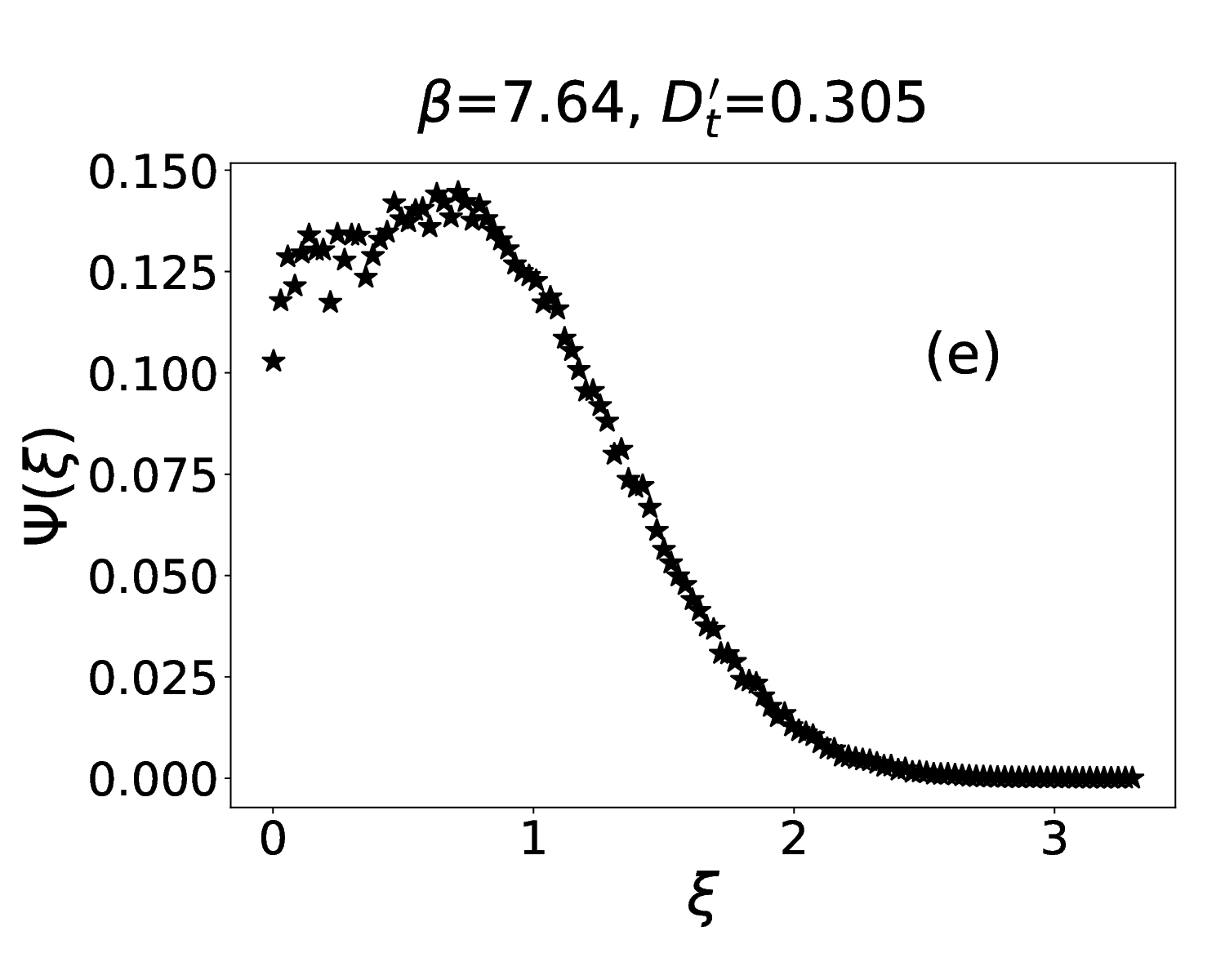}
  %\label{fig:plot_var0dt0}
\end{minipage} % <-- added
\vspace{0.5pt}
  \begin{minipage}{0.33\textwidth}
  \includegraphics[width=\linewidth]{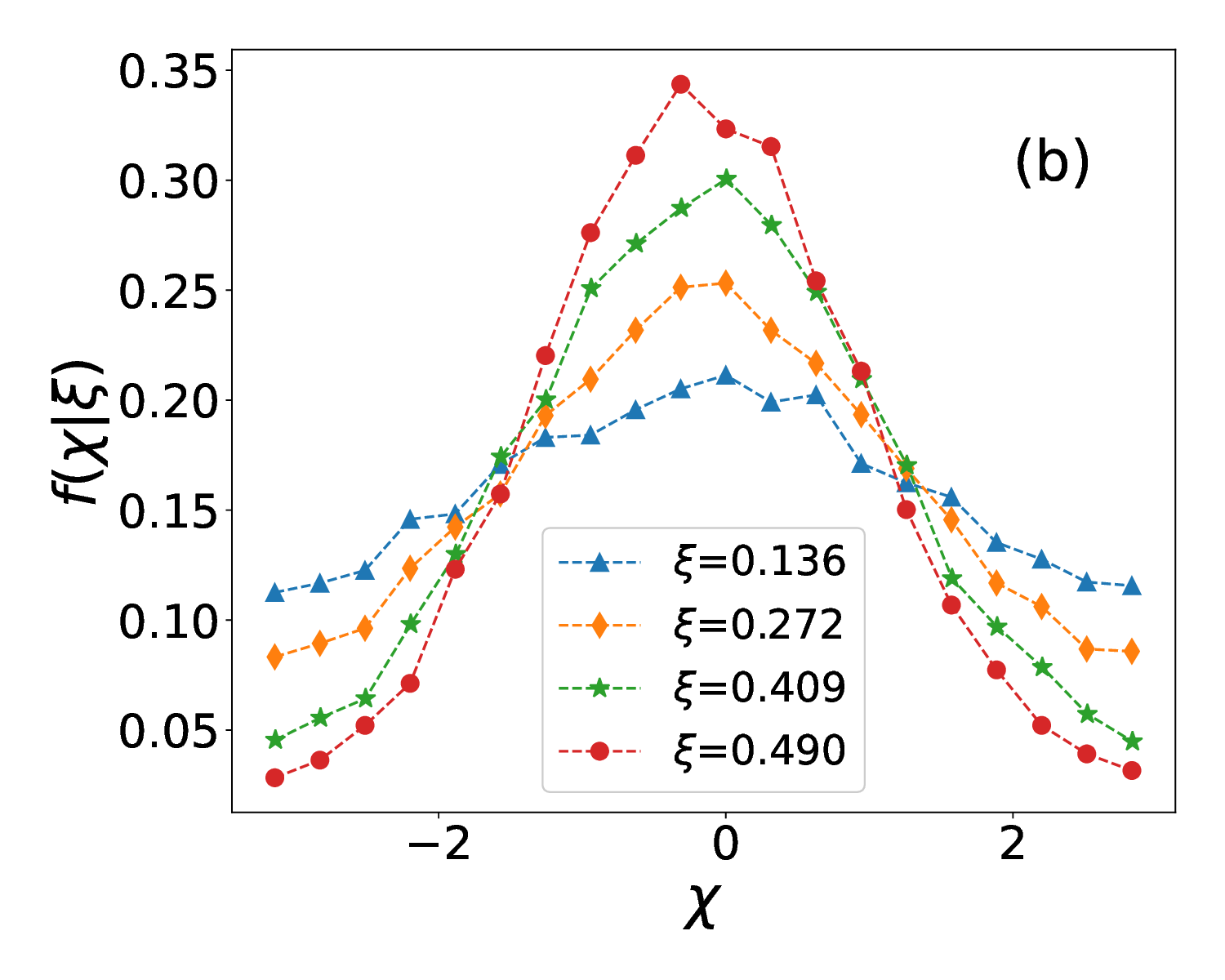}
  %\label{fig:plot_var2dt0}
\end{minipage} \hspace{0.01pt}% <-- added
\begin{minipage}{0.33\textwidth}
  \includegraphics[width=\linewidth]{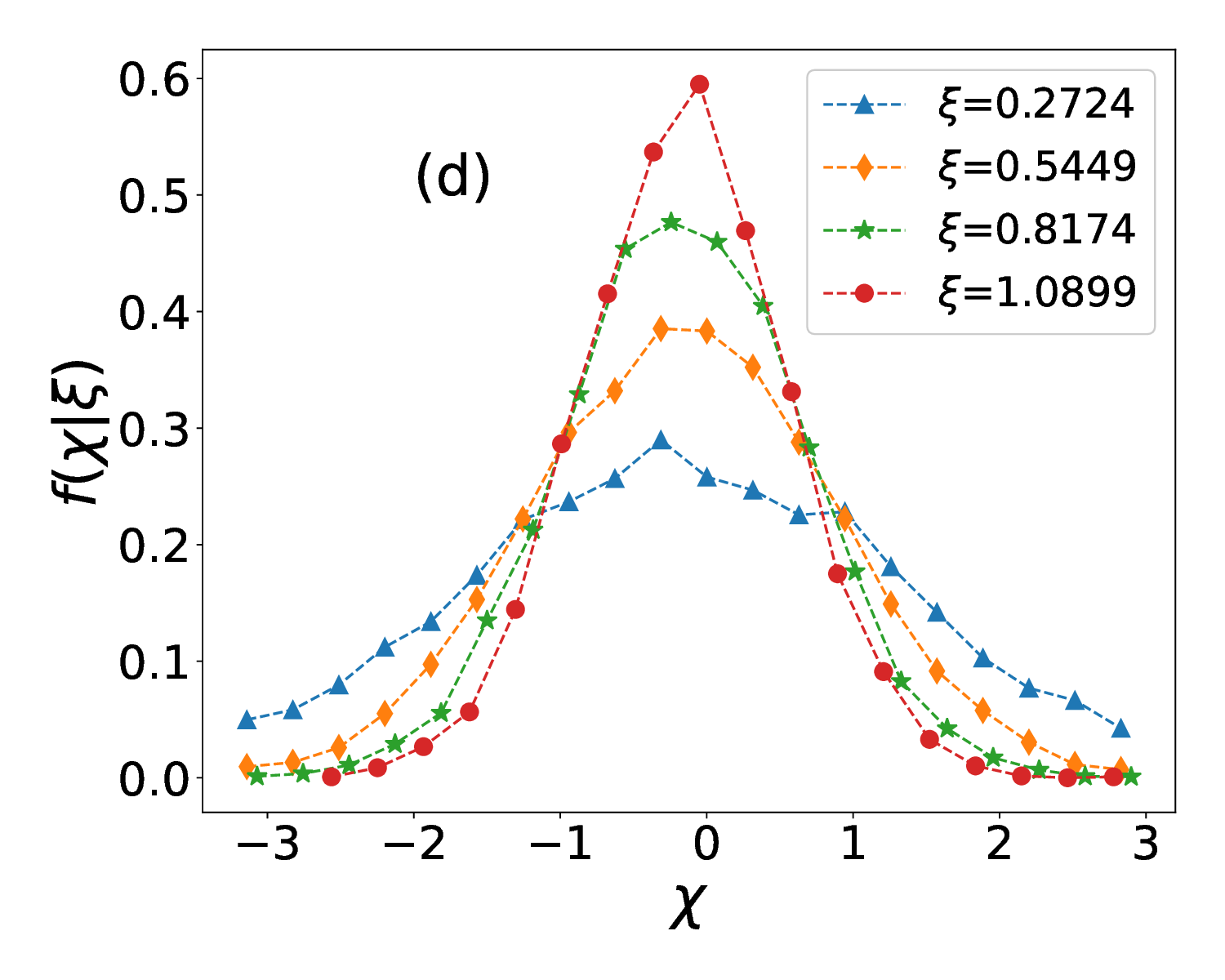}
  %\label{fig:plot_var1dt0}
\end{minipage}\hspace{0.01pt}% <-- added
\begin{minipage}{0.33\textwidth}
  \includegraphics[width=\linewidth]{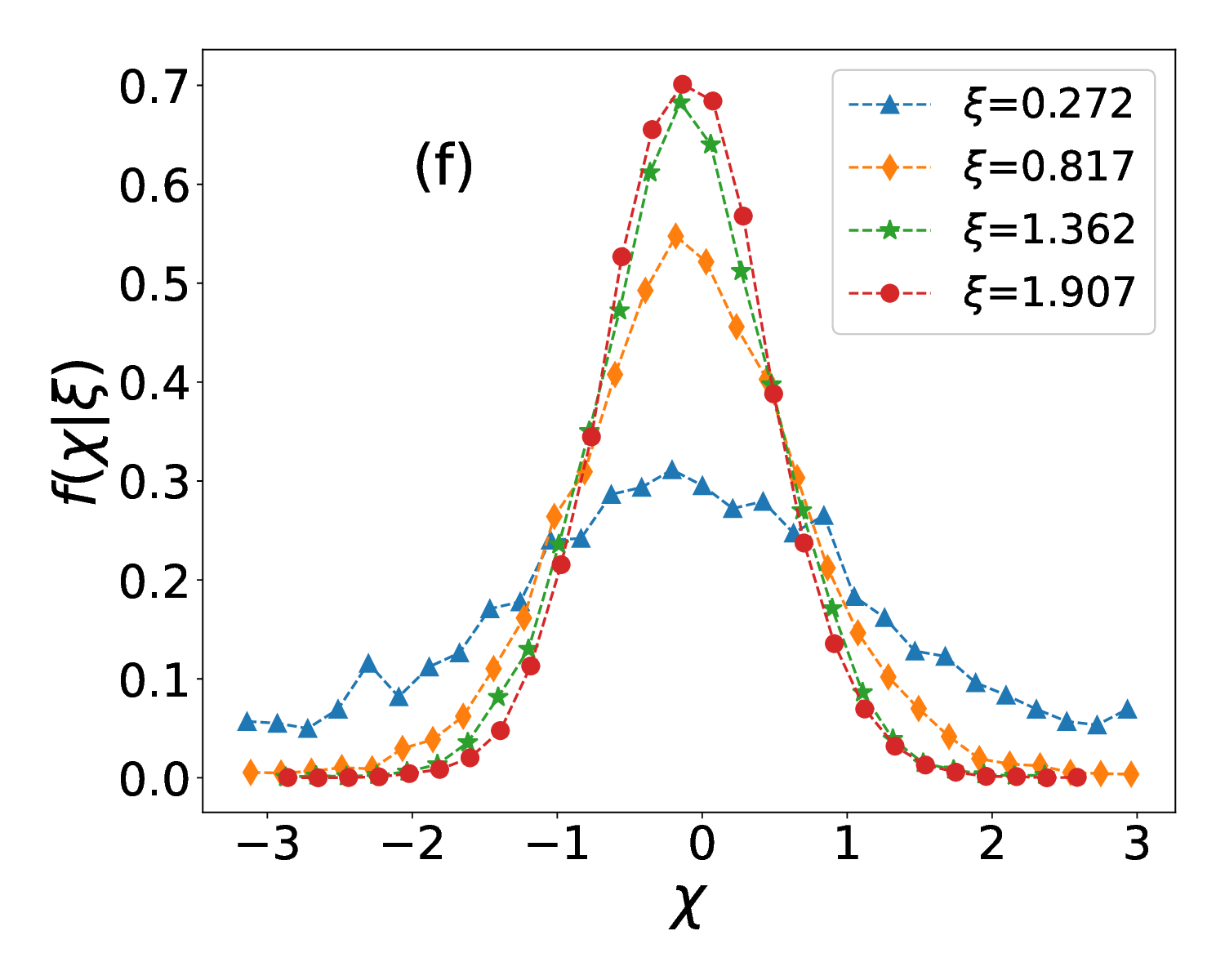}
  %\label{fig:plot_var0dt0}
\end{minipage} % <-- added
\caption{For non-zero translational diffusion coefficient $D_t=0.4$ $\mu$m$^{2}$s$^{-1}$, the figures show the radial distribution $\Psi(\xi)$ and angular distribution $f(\chi|\xi)$ for three different values of $\beta$:  (a,b) $\beta=0.076$, (c,d) $\beta=0.76$ and (e,f) $\beta=7.64$, respectively. The corresponding values of $\kappa$, for fixed $D_r$, are given in Table \ref{tab2}. The fixed parameter values can be found in Table~\ref{tab1}.}\label{fig:coschi_dt}
\end{figure*}

\begin{figure*}[htp]
    \centering % <-- added
    \begin{minipage}{0.45\textwidth}
  \includegraphics[width=\linewidth]{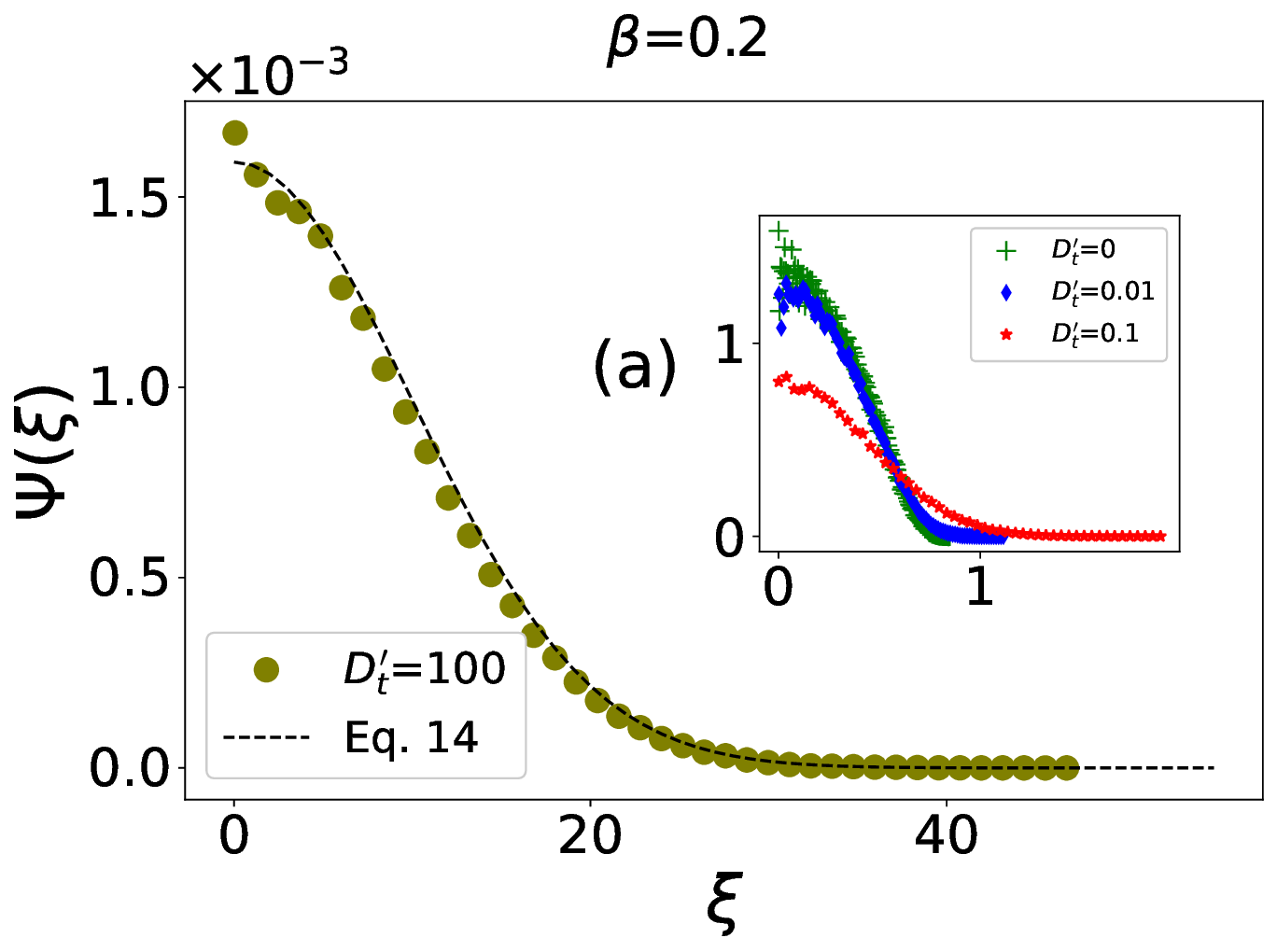}
  %\label{fig:plot_var2dt0}
\end{minipage} \hspace{0.01pt}% <-- added
\begin{minipage}{0.45\textwidth}
  \includegraphics[width=\linewidth]{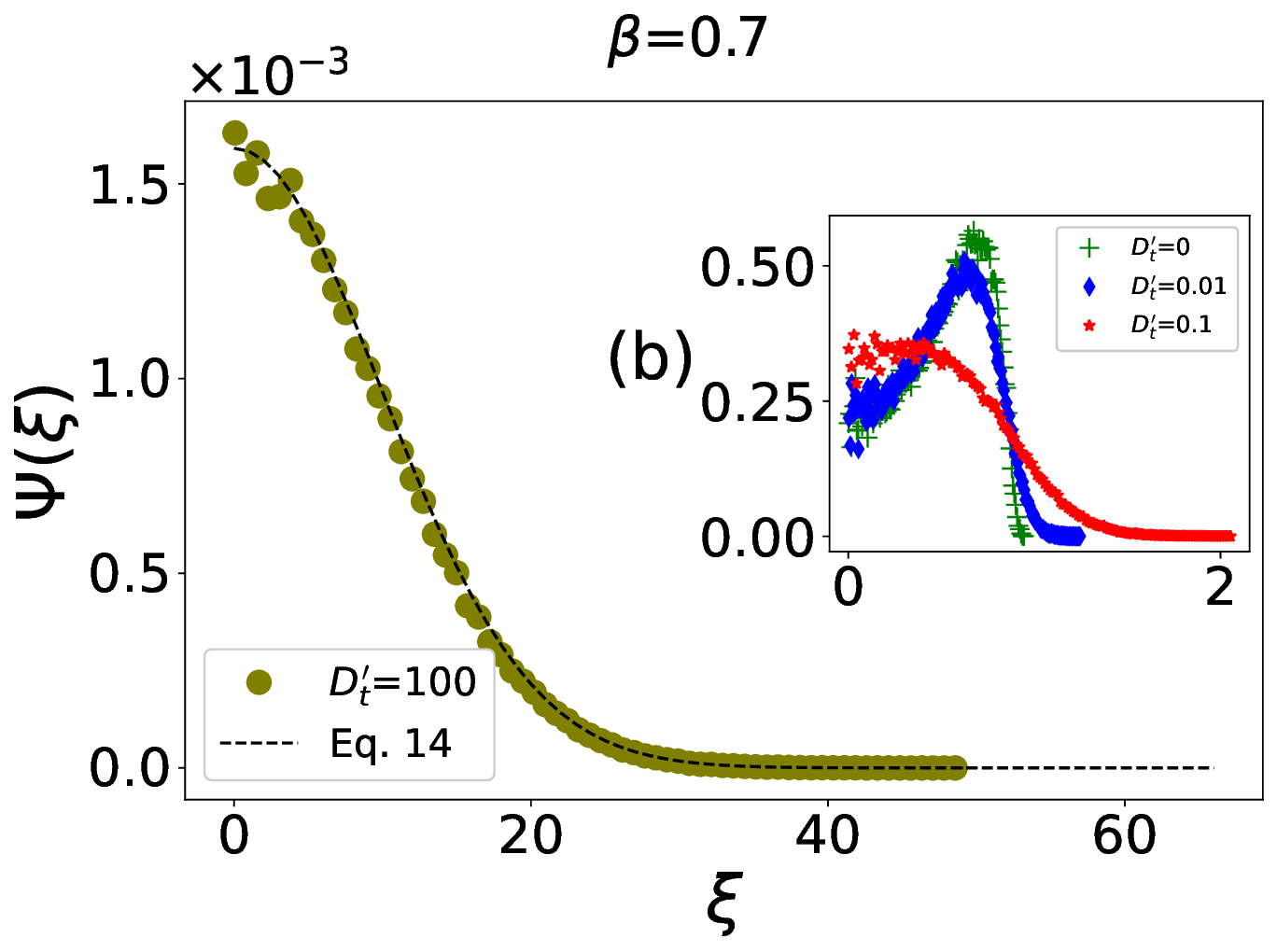}
  %\label{fig:plot_var1dt0}
\end{minipage}\hspace{0.01pt}% <-- added
\vspace{0.5pt}
\begin{minipage}{0.45\textwidth}
  \includegraphics[width=\linewidth]{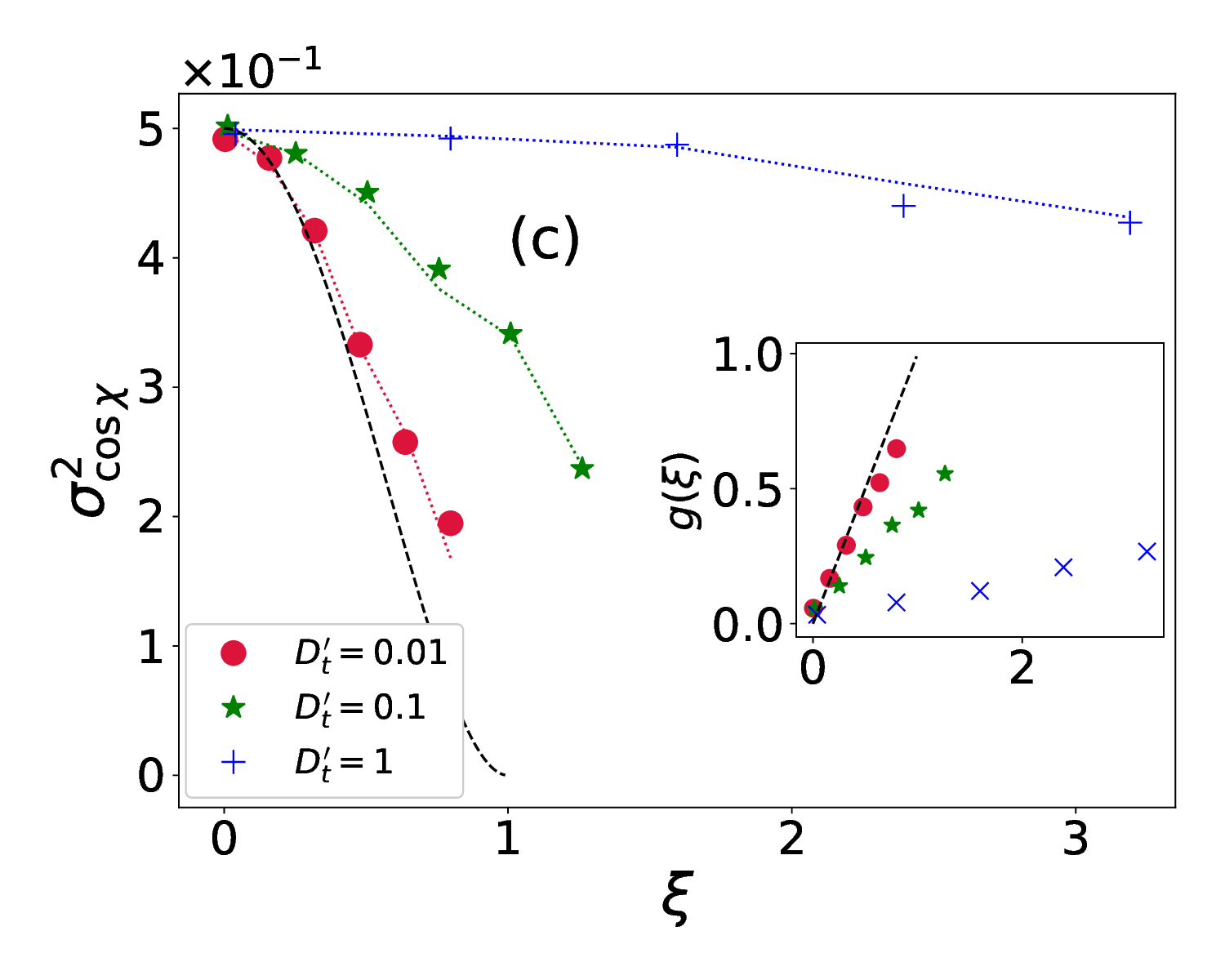}
  %\label{fig:plot_var0dt0}
\end{minipage} % <-- added
  \begin{minipage}{0.45\textwidth}
  \includegraphics[width=\linewidth]{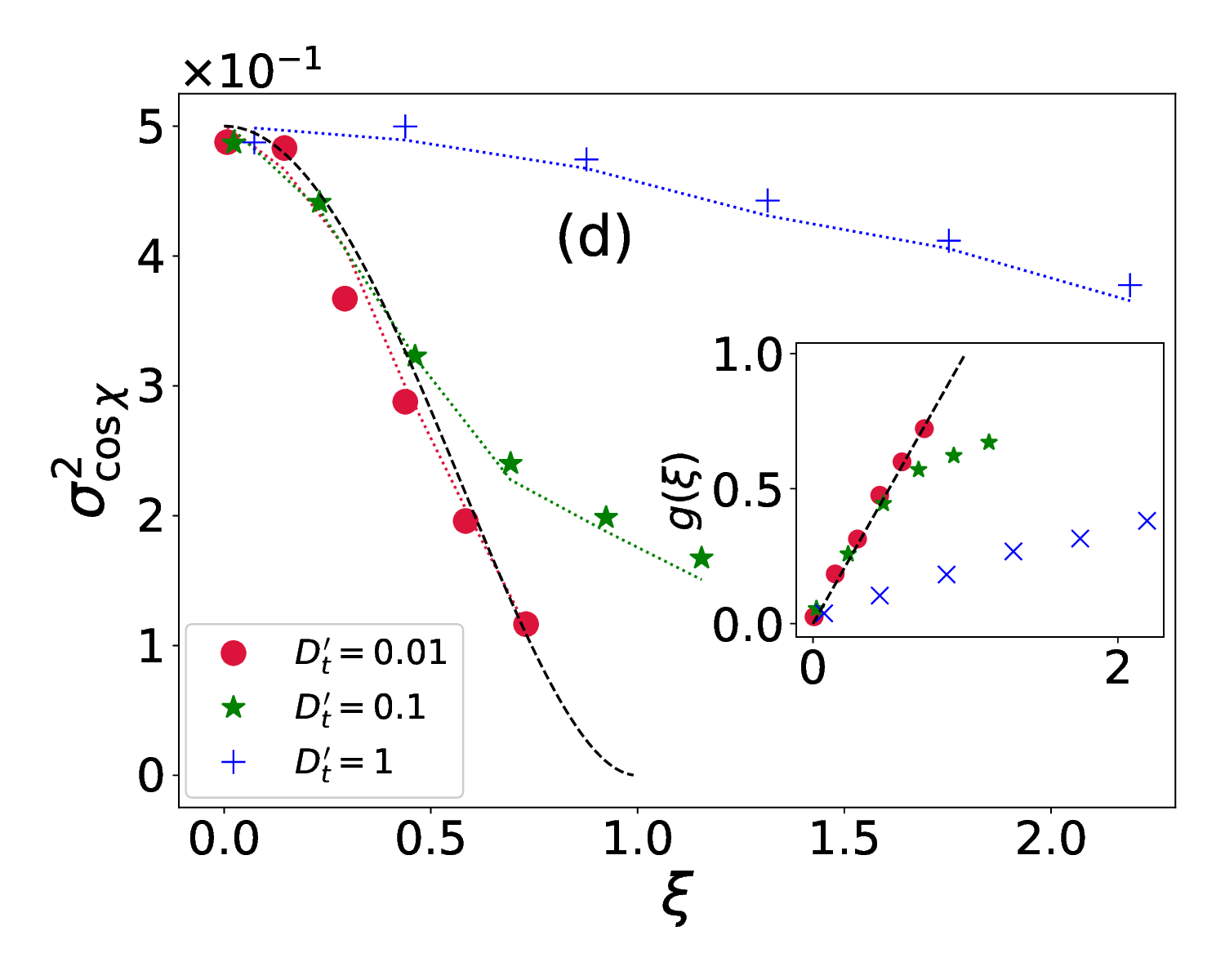}
  %\label{fig:plot_var2dt0}
\end{minipage} \hspace{0.01pt}% <-- added

\caption{The figures show that for fixed $\beta$, and increasing $D_t^{\prime}$, the positional distribution smoothly approaches the equilibrium Gaussian form. The simulation results are shown for two sets of parameters: (a,c): $\beta=0.2$ and (b,d): $\beta=0.7$, with $D^{\prime}_t$ taking different values as indicated in each figure (with $D_t$ fixed and $u_0$ varying). In (a) and (b) (main), the data is fitted with the equilibrium Gaussian distribution (Eq.~\ref{eq:psi}, with $g(\xi) \equiv 0$), and $T$ replaced by an effective temperature $T_{\rm eff}$. The best-fit values are given by (a) $T_{\rm eff}=$ 297.8 K and (b) $T_{\rm eff}$= 309.5 K  respectively. The insets show results for smaller values of $D_t^{\prime}$, including $D_t^{\prime}=0$ (for which we used $D_t=0$ and $u_0>0$). The figures (c) and (d) are results for mean(inset) and variance of $\cos{\chi}$ from simulations(symbols), fitted with the  theoretical expression obtained under the Gaussian assumption (Eq.~\ref{eq:app_var}) (lines), with $g(\xi)$ obtained from simulations. Dashed lines(black) in the main and inset figures represent the theoretical curve from Eq.~\ref{eq:u_bar} and Eq.~\ref{eq:variance_cos} at $D_t{'}=0$ for corresponding values of $\beta$.}
\label{fig:phi_dt_2b}
\end{figure*}

We also confirmed separately the predictions for the mean (Eq.~\ref{eq:u_bar})(inset) and the variance (Eq.~\ref{eq:variance_cos}) of $\cos{\chi}$, which are crucial for the derivation of the analytical result for $\Psi(\xi)$ (Eq.~\ref{eq:Phi}). In this case too, the theoretical predictions agree well with the numerical simulation results;  see Fig.~\ref{fig:coschi_dt0} (a,c and e). Interestingly, the angular distribution $f(\chi|\xi)$ shows a transition from unimodal to bimodal forms with an increase in trap stiffness, see Fig.~\ref{fig:coschi_dt0} (b,d and f), which coincides with the transformation in $\Psi(\xi)$ from concave to convex forms. In the unimodal regime, $\chi$ is centered around zero, suggesting that the motion of the particles is, predominantly radial, oriented outwards from the center. In the bimodal regime, on the other hand, the motion is preferentially tangential, with the particles orbiting around the center of the trap equally in clockwise and counterclockwise modes. Our Gaussian approximation for $\chi$, used in deriving Eq.~\ref{eq:variance_cos} for the variance in $\cos\chi$, however, does not take into account this transition from unimodal to bimodal forms. This limitation is reflected in the discrepancy between simulation results and analytic prediction for very large $\kappa$ (see Fig.~\ref{fig:coschi_dt0}, part (c)). 

Experimentally, a similar transition in the motion of active particles has been reported in a recent paper by Dauchot and D\'{e}mery\cite{dauchot2019dynamics}. Here, the authors studied the motion of a hexbug nano (a toy robot with dimensions 45 mm$\times$ 15 mm$\times$ 15 mm), moving on a parabolic dish. As the propulsion velocity of the hexbug was decreased, it was observed to undergo a  transition from ``orbiting motion" to ``climbing motion". In the orbiting regime, the hexbug rotates around the bottom of the parabola, with fluctuations in its radial position. In our study, we observe a transition to orbiting motion (characterized by symmetrically located off-center peaks in the angular probability distribution), as the trap stiffness is increased, keeping other parameters fixed (see  Fig~\ref{fig:vector_plot_2d}b and \ref{fig:coschi_dt0}f). Indeed, from Eq.~\ref{eq:dimensionlessparameter}, we see that reducing $\kappa$ at fixed $u_0$ is equivalent to increasing $u_0$ at fixed $\kappa$; in both cases, the system is pushed towards the active (orbiting) regime. In the passive (climbing) regime, the hexbug predominantly points radially away from the center, with unimodal angular distribution (see  Fig~\ref{fig:vector_plot_2d}a and \ref{fig:coschi_dt0}b). Our simulation results are consistent with the experimental observations in \cite{dauchot2019dynamics}.

\subsection{$D_t^{\prime}>0$: active to passive crossover}
We will now discuss our results for simulations with non-zero $D_t$. Unless otherwise specified, the simulation parameters mentioned in Table~\ref{tab1}. 

Unlike the $D_t=0$ case discussed in the previous subsection, where the positional distribution shows a single transition between concave and convex forms, simulation data for non-zero, fixed $D_t$ shows slightly different behavior of $\Phi(r)$ with the increase in $\kappa$ (keeping other parameters fixed), see Fig.~\ref{fig:coschi_dt}(a,c and e). Here, for small $\beta$ = $0.076$, $\Psi(\xi)$ is concave, with peak at the origin, similar to Fig.~\ref{fig:phi_dt}(a). For intermediate $\beta$ = $0.76$, the peak has moved away from the origin and the particle density becomes a non-monotonic function of the distance from the origin, similar to the corresponding $D_t=0$ result in Fig.~\ref{fig:phi_dt}(inset of (b)). For large $\beta$ = $7.64$, however, the peak moves closer to the origin, approaching a form that is superficially similar to the concave curve in the small $\kappa$ limit(see, Fig.~\ref{fig:phi_dt}(b)). The possible existence of a re-entrant transition was first reported in the theoretical study by Malakar et al.~\cite{malakar2020steady}.

Does the convex-concave shift between intermediate and large $\kappa$ values indicate a re-entrant transition? We believe that the answer is more subtle. A comparison of the values of the dimensionless parameters $\beta$ and $D_t^{\prime}$ is instructive. For non-zero $D_t$, increasing $\kappa$ at fixed $u_0$ increases $\beta$ as well as $D_t^{\prime}$. While the increase of $\beta$ tends to move the peak away from the origin, a simultaneous increase in $D_t^{\prime}$ moves the system towards the passive, equilibrium behavior. Therefore, we believe that the distribution in Fig.~\ref{fig:coschi_dt}(a) is a slightly modified form of the one in Fig.~\ref{fig:phi_dt}(a), while the one in Fig.~\ref{fig:coschi_dt}(e) is close to the equilibrium, Gaussian form.

For the angular distribution $f(\chi|\xi)$, our simulation results are shown in Fig.~\ref{fig:coschi_dt} (b,d and f), for three different values of $\beta$. Here, we observe that the orientation remains predominantly radial at all values of $\beta$, irrespective of the distance from the trap center. The transition between unimodal and bimodal forms, observed for $D_t^{\prime}=0$, appears to be absent here. 
\begin{table}[t]
\begin{tabular*}{0.48\textwidth}{@{\extracolsep{\fill}}llllll}
\hline
 $k_B T$ &$a$ & $D_t$ & $D_r$ &$u_0$ \\ 
 (pN nm) &($\mu$m) & ($\mu$m$^2$ s$^{-1})$ & (${\rm rad}^2$ s$^{-1})$ & ($\mu$m $s^{-1})$ \\  
 \hline
 4.11 & 0.53 & 0.4 & 1.308 & 3.67    \\
 \hline
\end{tabular*}
\caption{A list of the standard parameters used in our numerical simulations. The value of $k_B T$ corresponds to temperature $T = 300 K$. For an explanation of the reasons behind the chosen values, please see the text.}
 \label{tab1}
\end{table}
In order to explore in more depth the transition between active and passive regimes, we also carried out simulations where $u_0$ was reduced from its ``standard" value in Table ~\ref{tab1}, and thereby $D_t^{\prime}$ increased. The results for the position distribution $\Psi(\xi)$ are shown for two values of $\beta$, 0.2 and 0.7 in Fig.~\ref{fig:phi_dt_2b} (a and b, respectively). For fixed $\beta$ and $D_t$, as $D_t^{\prime}$ is increased above zero by reducing $u_0$, we observe that $\Psi(\xi)$ approaches the equilibrium Boltzmann distribution, as expected from Eq.~\ref{eq:formalsolution} (see insets). For $D_t^{\prime}\gg 1$, the non-scaled distribution $\Phi(r)$ fitted well with a Gaussian curve $\Phi(r)\propto \exp(-\kappa r^2/2k_B T_{\rm eff})$, with $T_{\rm eff}\to T$, the equilibrium temperature, as $u_0\to 0$. More detailed  discussions are given in the caption of Fig.~\ref{fig:phi_dt_2b}.    

\begin{table}[t]
\begin{tabular*}{0.2\textwidth}{@{\extracolsep{\fill}}|l|l|}
\hline
$\beta$ & $\kappa$ (pN nm$^{-1}$) \\ 
 \hline
0.076 & $10^{-6}$  \\
 0.76 & $10^{-5}$    \\
7.64 & $10^{-4}$   \\
 
 \hline
\end{tabular*}
\caption{The most commonly used values for $\kappa$ in our simulations, along with the corresponding values of the dimensionless parameter $\beta$ (Eq.~\ref{eq:beta}), when $D_r$ is held fixed (see Table ~\ref{tab1}).}
 \label{tab2}
\end{table}
We also studied the $\xi$-dependence of mean and variance of $\cos\chi$ in the above simulations. In Fig.~\ref{fig:phi_dt_2b}, we show our results for two values of $\beta$, 0.2 and 0.7 (c and d, respectively). These values are chosen such that they fall on either side of the critical value ($\beta_c=1/3$ in the Gaussian $\chi$-approximation). In both cases, $g(\xi)$ is an increasing function of $\xi$ for all values of $D_t^{\prime}$, with $g(\xi)\to 0$ as $\xi\to 0$. The variance $\sigma^2_{\cos\chi}$, on the other hand, is a decreasing function of $\xi$ in all cases. Also, the simulation data fits well with the Gaussian expression given in Eq.~\ref{eq:app_var}, when $g(\xi)$ is extracted from simulation data. 
\subsection{A ``phase diagram'' for the ABP in $d=2$}
\begin{figure*}
\centering
    \begin{minipage}{0.5\textwidth}
  \includegraphics[width=\linewidth]{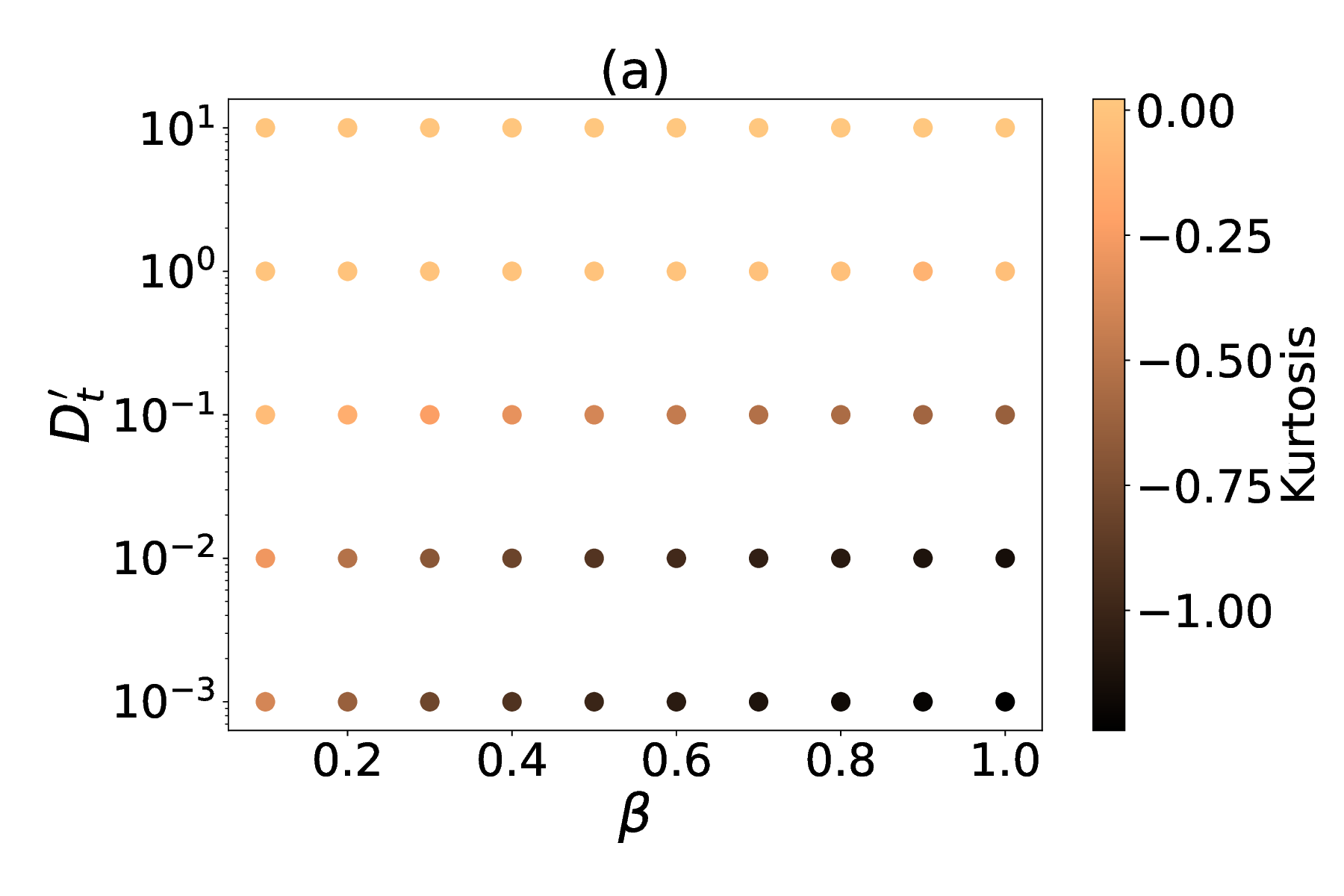}
\end{minipage} 
\begin{minipage}{0.40\textwidth}
  \includegraphics[width=\linewidth]{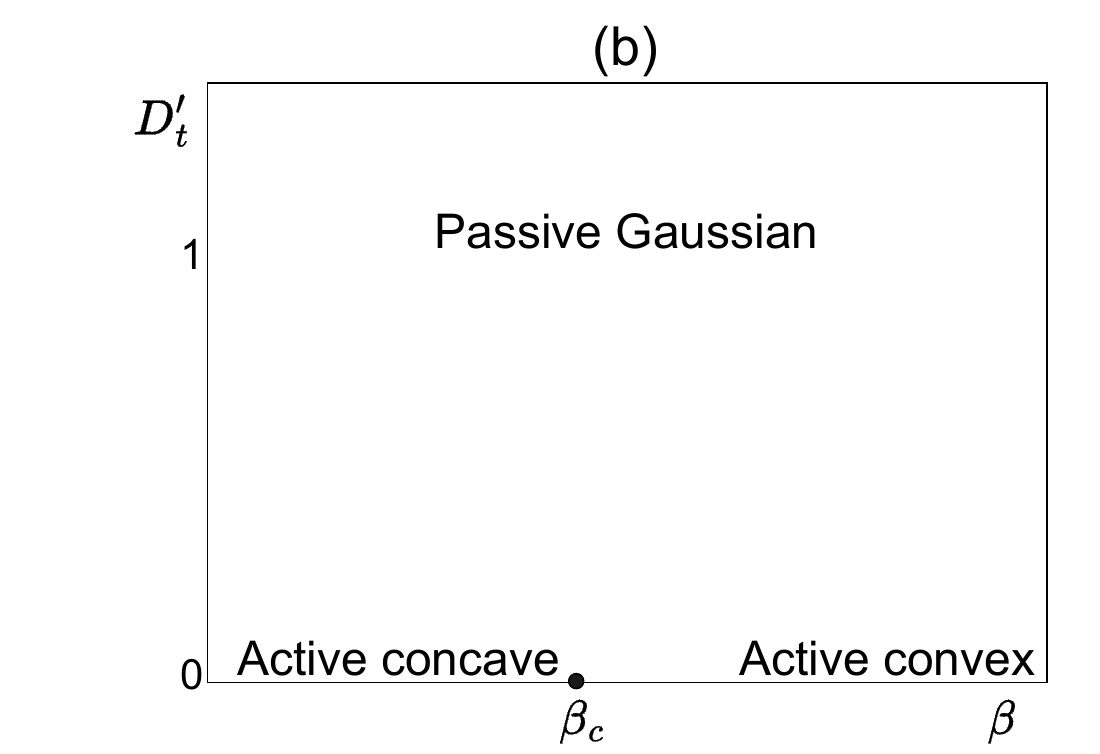}
\end{minipage}% <-- added
\caption{(a) A heat map for the kurtosis of the marginal distribution $p(x)$ (Eq.~\ref{eq:marginal}), obtained from simulations, is shown in (a), with $D^{\prime}_t$ and $\beta$ as independent parameters. In order to change $D_t^{\prime}$, we changed $u_0$ and to change $\beta$, $\kappa$ was changed with $u_0$ kept fixed. All other parameters have fixed values as given in Table~\ref{tab1}. (b) Based on our theoretical understanding, simulation results and the observations in (a), we constructed a schematic ``phase diagram" to represent different shapes for the positional distribution. These are identified as active concave, active convex (both along and close to the $D_t^{\prime}=0$ line) and passive Gaussian (for $D_t^{\prime}\gg 1$). Along the $D_t^{\prime}=0$ line, the active concave and convex shapes are separated by a ``critical point", denoted $\beta=\beta_c$, with $\beta_c\approx 1/3$.}
\label{fig:phase_daigram}
\end{figure*}

In order to construct a complete picture capturing the transitions between three different regimes, i.e., concave, convex and equilibrium, we have tried to construct a ``phase diagram" for the positional distribution of the ABP in a harmonic trap, with the dimensionless parameters $\beta$ and $D_t^{\prime}$ on the two axes. To vary $\beta$, we varied $\kappa$ in the range $10^{-6}-10^{-4}$ pN nm$^{-1}$, with $D_r$ kept fixed. To vary $D_t^{\prime}$, we varied $u_0$ in the range 100$\mu$m s$^{-1}-$24 mm s$^{-1}$ and $\kappa$, with $D_t$ held fixed. The other parameters were kept fixed at their standard values (Table~\ref{tab1}). 
The deviation of the marginal distribution 
\begin{equation}
    p(x)=\int\Phi(x,y)dy
    \label{eq:marginal}
\end{equation}
from Gaussianity can be measured by its kurtosis. Using our simulation data, we constructed a kurtosis heat map~\cite{chaudhuri2021active,caprini2022parental} for the system in Fig.~\ref{fig:phase_daigram}(a), with $D_t^{\prime}$ and $\beta$ as independent parameters. With increase in $D_t^{\prime}$, at fixed $\beta$, the distribution approaches the equilibrium Boltzmann form, which is Gaussian. On the other hand, as $\beta\to 0$, for fixed $D_t^{\prime}$, we approach yet another Gaussian limiting form, but this time, the ``active" Gaussian. This active Gaussian is an asymptotic limiting form of a more general, active concave distribution, which is separated from the active convex distribution by a ``critical point", $\beta=\beta_c$, along the $D_t^{\prime}=0$ line. Both the active concave distributions (concave and convex) are expected to smoothly cross over to the equilibrium Gaussian (Boltzmann) distribution with increase in $D_t^{\prime}$.

In a schematic phase diagram shown in Fig.~\ref{fig:phase_daigram}(b), we have roughly identified the locations of the different ``phases" as below: 

 (a) Active concave phase ($D_t^{\prime}=0$): $\Psi(\xi)$ has maximum at the trap centre ($\xi=0$), but is, in general, non-Gaussian. In the limit $\beta\to 0$, this distribution tends to the active Gaussian form (Eq.~\ref{eq:active_gaussian}).

(b) Active convex phase ($D_t^{\prime}=0$): $\Psi(\xi)$ has a maximum away from $\xi=0$, and either a minimum at $\xi=0$. 

(c) Passive, equilibrium phase ($D_t^{\prime}\gg 1$): For $D_t^{\prime}>0$, for $\beta<\beta_c$, the distribution locally resembles the equilibrium Gaussian form close to $\xi=0$. For $\beta>\beta_c$, it acquires a secondary maximum at the origin, in addition to the primary maximum away from it. With further increase in $D_t^{\prime}$, the primary maximum loses height, moves closer to the origin and eventually merges with the maximum at $\xi=0$ which becomes dominant for large $D_t^{\prime}$(inset of Fig.~\ref{fig:phi_dt_2b} b). In this limit, $\Psi(\xi)$ is nearly Gaussian in shape, and very closely approximates the equilibrium, Boltzmann form. As $D_t^{\prime}\to \infty$, at fixed $\beta$, the distribution approaches the Boltzmann distribution.

In general, the results compiled together in Fig.~\ref{fig:phase_daigram} agree with our expectations based on scaling arguments and explicit mathematical results presented in Sec.II. 
\subsection{Harmonically confined ABP in $d=3$}
We also carried out a limited number of simulations in three dimensions, where the unit propulsion vector $\hat{\bf u}=(\sin{\theta_u}\cos{\phi_u},\sin{\theta_u}\sin{\phi_u},\cos{\theta_u})$,  $\theta_u$ and $\phi_u$ being the polar and azimuthal angles, respectively. The harmonic trapping potential too becomes three-dimensional, with potential energy $U({\bf r})=(\kappa/2)r^2$ where $r^2=x^2+y^2+z^2$.

\begin{figure*}[t]
    \centering % <-- added
    \begin{minipage}{0.45\textwidth}
  \includegraphics[width=\linewidth]{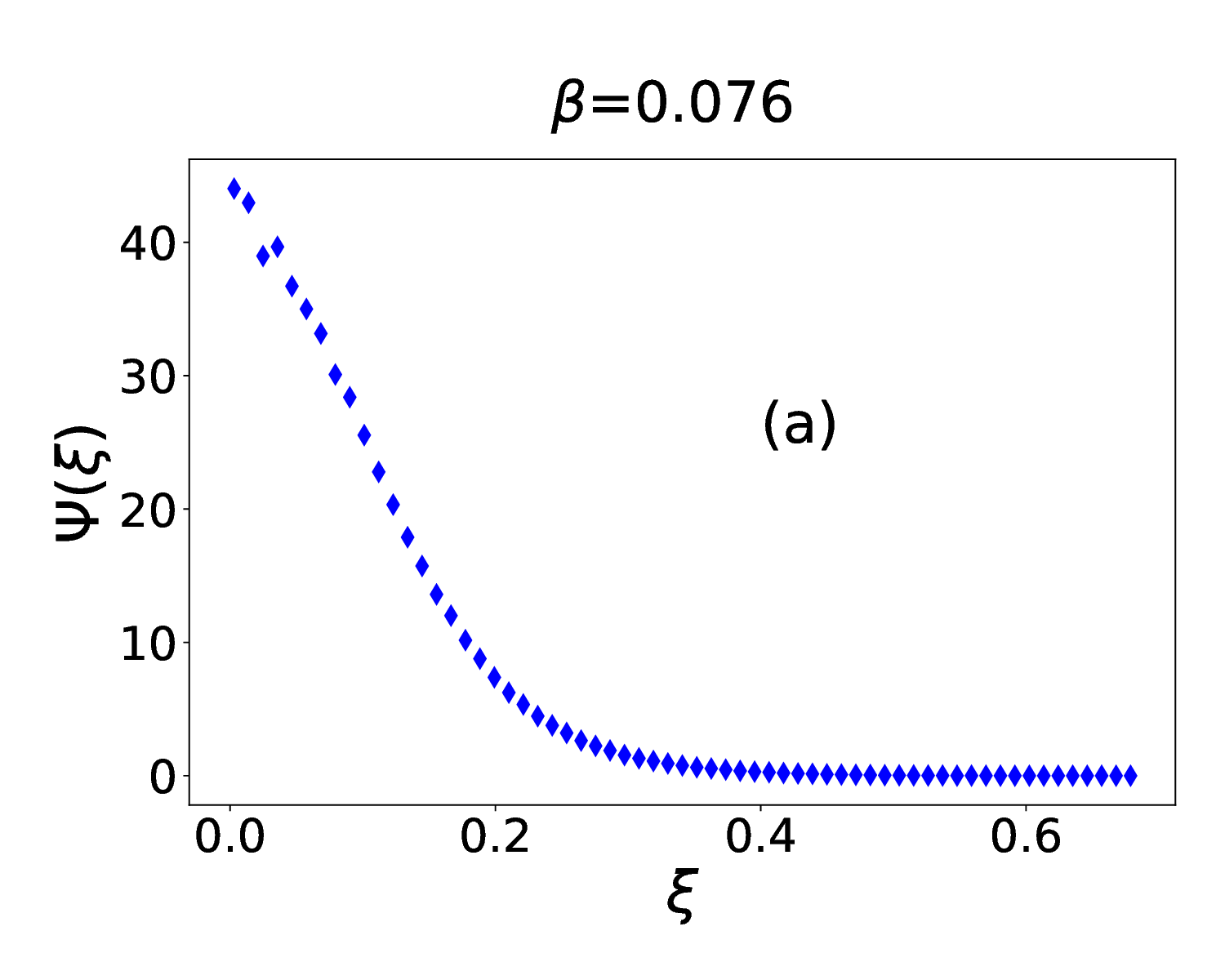}
  %\label{fig:plot_var2dt0}
\end{minipage} \hspace{0.01pt}% <-- added
\begin{minipage}{0.45\textwidth}
  \includegraphics[width=\linewidth]{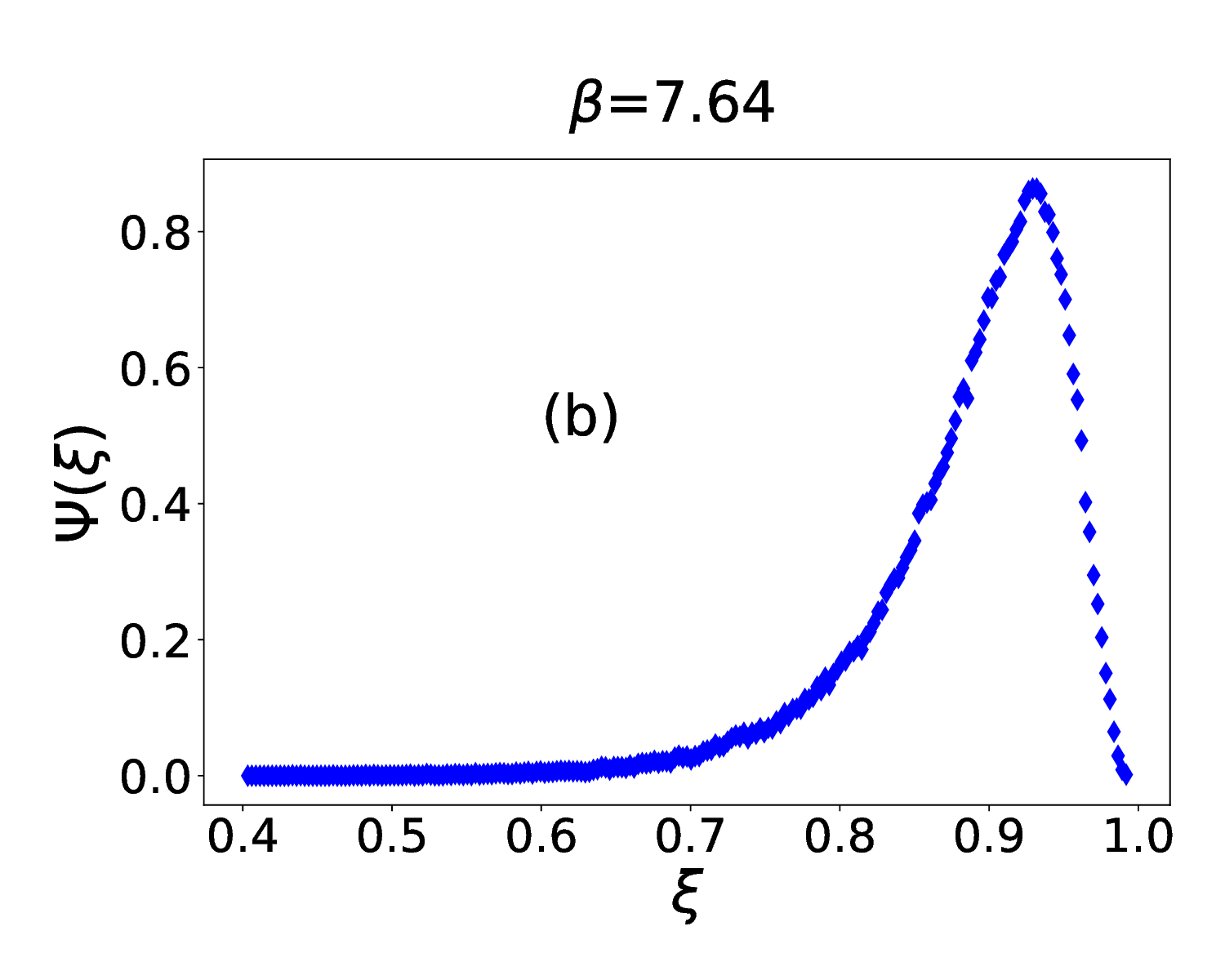}
  %\label{fig:plot_var1dt0}
\end{minipage}\hspace{0.01pt}% <-- added
\vspace{0.5pt}
\begin{minipage}{0.45\textwidth}
  \includegraphics[width=\linewidth]{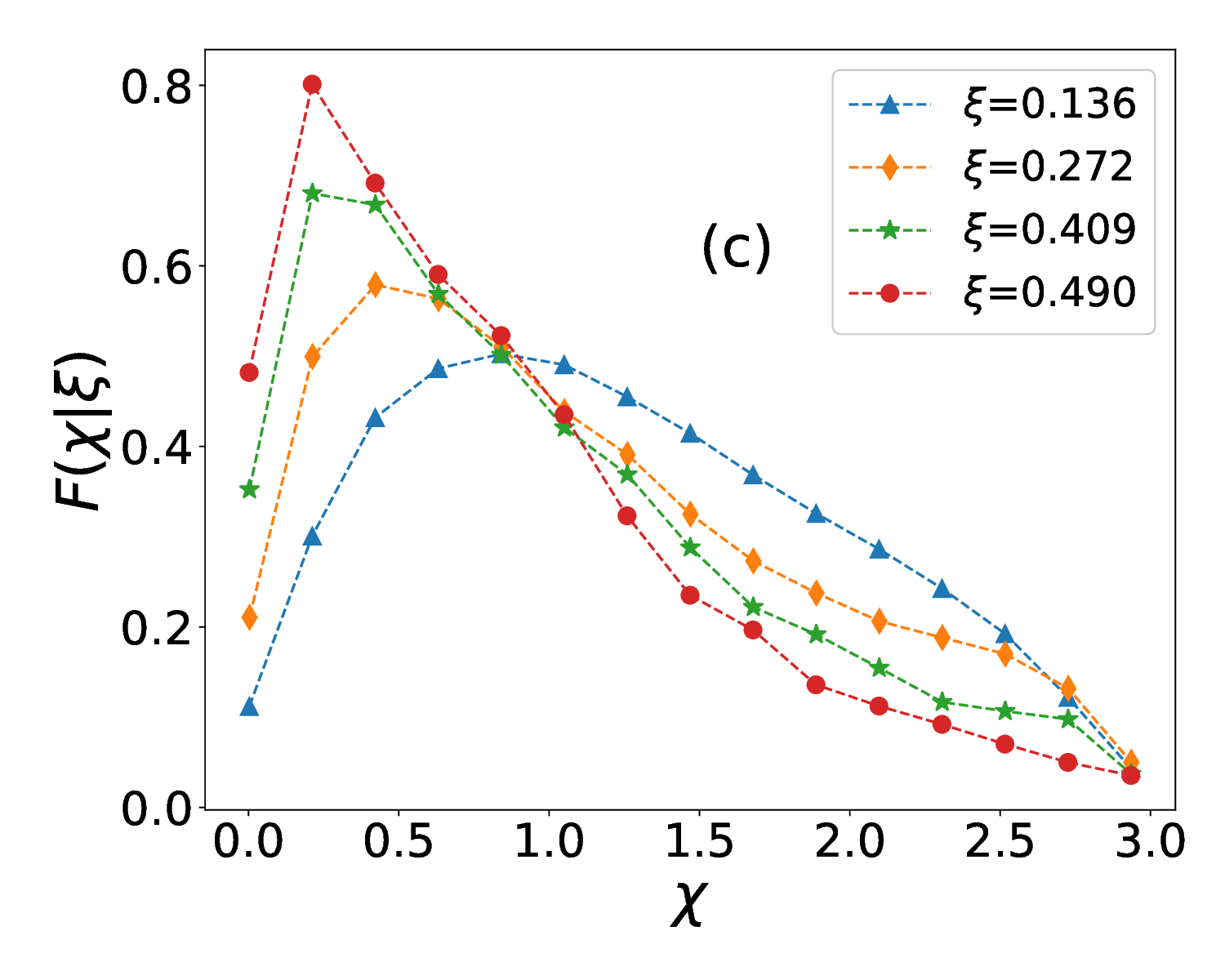}
  %\label{fig:plot_var0dt0}
\end{minipage} % <-- added
  \begin{minipage}{0.45\textwidth}
  \includegraphics[width=\linewidth]{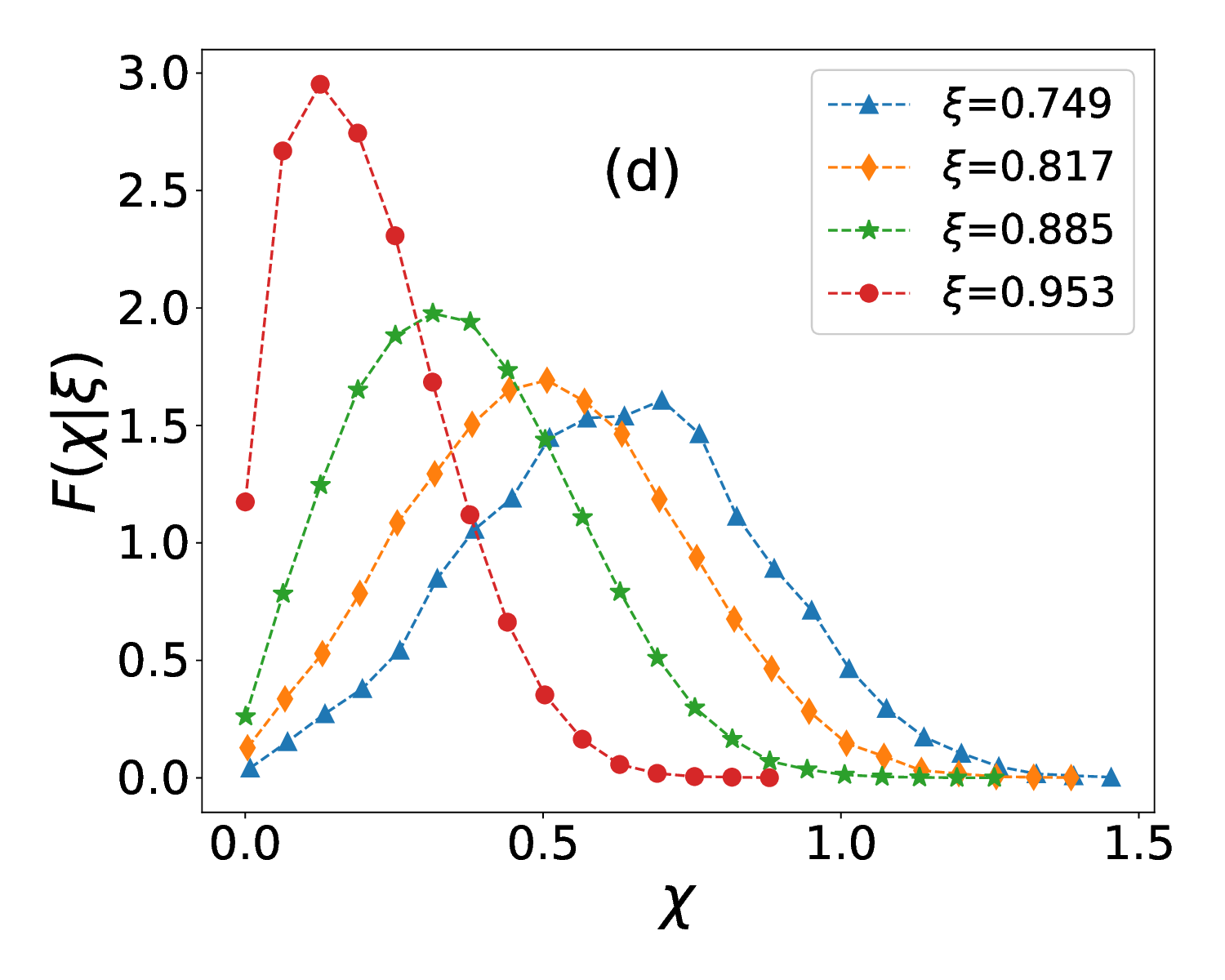}
  %\label{fig:plot_var2dt0}
\end{minipage} \hspace{0.01pt}% <-- added

\caption{In three dimensions, for $D_t = 0$ and $u_0>0$, we have shown simulation results in first row for $\Psi(\xi)$ and second row shows for angular distribution  $F(\chi|\xi)$, for particular values of $\beta$,  (a,c) 0.076 and (b,d) 7.64 respectively. The corresponding values of $\kappa$, with $D_r$ fixed, are given in Table \ref{tab2}. The  numerical values of the fixed parameters (such as $D_r$ and $u_0$) are given in Table~\ref{tab1}. }
\label{fig:phi_3d}
\end{figure*}

The general form of the over-damped Langevin equations remains the same as in Eq.~\ref{eq:langevin_main}. Here, ${\boldsymbol \eta}_r\equiv (\eta_{rx}, \eta_{ry}, \eta_{rz})$, where $\eta_{rx}$, $\eta_{ry}$ and $\eta_{rz}$ are independent, uncorrelated Gaussian white noise terms with zero mean and unit variance. After substituting the expression for $\hat{\bf u}$ expressed in spherical polar coordinates, 
Eq.~\ref{eq:langevin_main} becomes~\cite{das2018confined}, in $d=3$, 
\begin{equation}
    \begin{aligned}
        & \dot{\theta}_u=\eta_{ry}~\cos{\phi_u}-\eta_{rx}~ \sin{\phi_u}\\
        & \dot{\phi}_u=\eta_{rz}-\cot{\theta_u}[\eta_{rx}~ \cos{\phi_u}+\eta_{ry}~ \sin{\phi_u}].
        \label{eq:3drotation}
    \end{aligned}
\end{equation}
Similar to $d=2$, the position and orientation of the ABP are updated by using the Euler forward discretization scheme, which gives the following equations: 

\begin{equation}
    \begin{aligned}
       & x_{i+1}=x_i +\Delta t [u_0 ~\sin{\theta_{u,i}}\cos{\phi_{u,i}}-k x_i],\\
        & y_{i+1}=y_i +\Delta t [u_0 ~\sin{\theta_{u,i}}\sin{\phi_{u,i}}-k y_i],\\
         & z_{i+1}=z_i +\Delta t [u_0 ~\cos{\theta_{u,i}}-k z_i],
    \end{aligned}
\end{equation}
and 
\begin{equation}
\begin{aligned}
&\theta_{u,i+1}= \\ &=\theta_{u,i}+\sqrt{2D_r \Delta t}~[\eta_{ry,i}~ \cos{\phi_{u,i}}-\eta_{rx,i} ~\sin{\phi_{u,i}}],
\end{aligned}
\end{equation}
\begin{equation}
\begin{aligned}
   & \phi_{u,i+1}=\phi_{u,i}+\sqrt{2D_r \Delta t}~\times \\ &\times [\eta_{rz,i}-\cot{\theta_{u,i}}(\eta_{rx,i}~\cos{\phi_{u,i}}+\eta_{ry,i}~\sin{\phi_{u,i}})]
    \end{aligned}
\end{equation}
In $d=3$, only the special case $D_t=0$ was studied by us in simulations. Some results for the positional distribution $\Psi(\xi)=r_m^{3}\Phi(r_m\xi)$ are shown in Fig.~\ref{fig:phi_3d}(a) and (b), where $\xi=r/r_m$, with $r_m$ defined in Eq.~\ref{eq:rm}. Similar to $d=2$, we observe that the distribution undergoes transition from concave to convex shapes with increase in $\beta$, with a strict cut-off at $\xi=1$.

\begin{figure}[h]
\includegraphics[width=\linewidth]{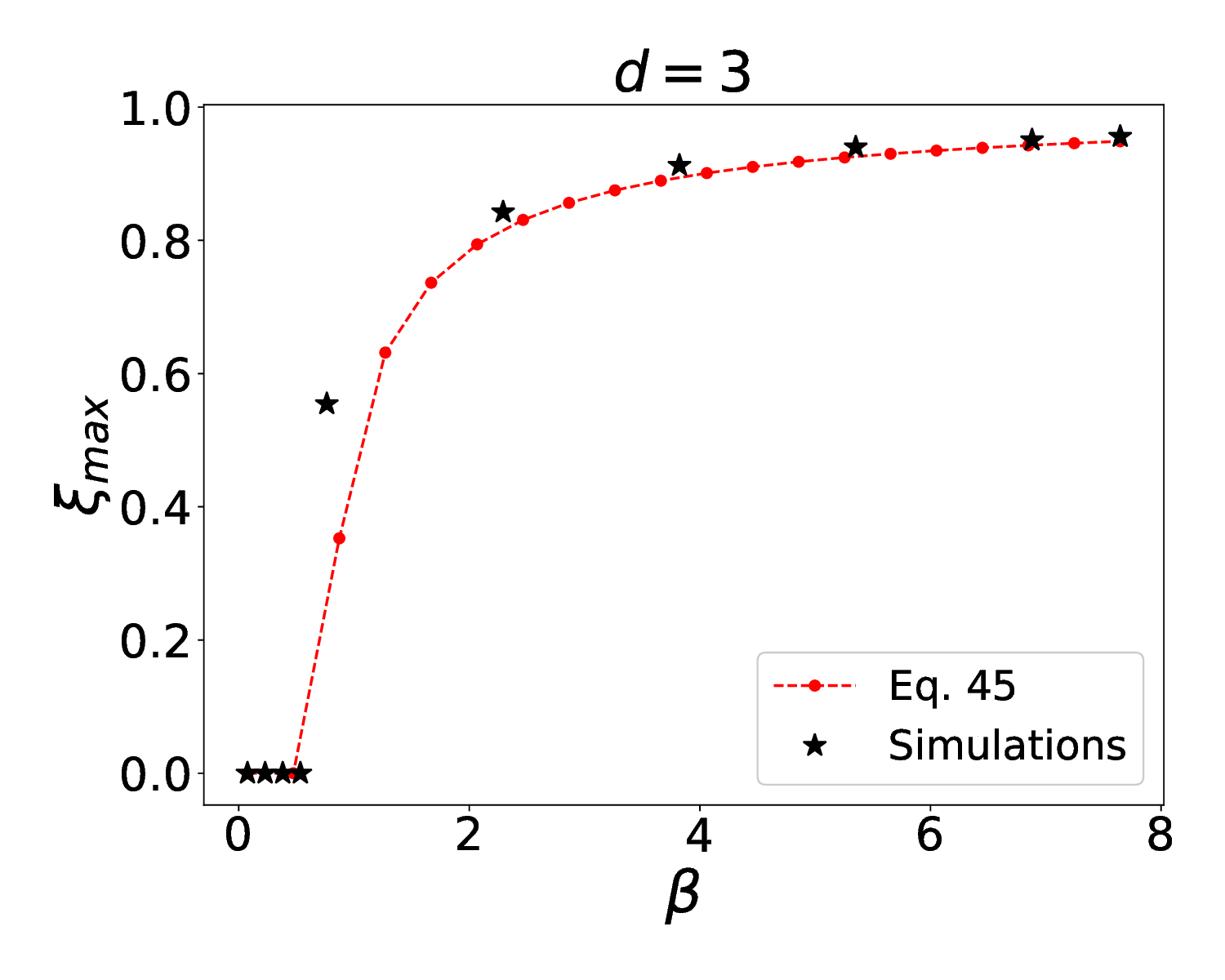}
\caption{In three dimensions, simulation results (black, diamond) for $\xi_{\rm max}$ shows a transition similar to $d=2$, but at a different critical value. The dashed line is a fit (Eq.~\ref{eq:rma}, red circles), with best fit parameters $\beta_c=0.535$ and $a=0.435$.} 
\label{fig:rmax_3d}
\end{figure}

To understand the orientational profile of the ABP in $d=3$, we defined the solid angle $\chi=\arccos(\hat{\bf r}\cdot \hat{\bf u})$, which characterises the orientation of the propulsion vector with respect to the outward radial direction ($0\leq \chi\leq \pi$). In simulations, the conditional probability distribution $F(\chi|\xi)$ for the angle $\chi$, at fixed $\xi$, is defined as 
\begin{equation}
  F(\chi|\xi)\equiv \frac{\int P(r_m{\boldsymbol \xi}, {\hat {\bf u}})\delta[\chi-\arccos(\hat{\bf r}\cdot \hat{\bf u})]d^3{\hat{\bf u}}}{\Phi(r_m\xi)}. 
    \label{eq:conditional3d}
\end{equation}
In Fig.~\ref{fig:phi_3d}(c) and (d), we plot $F(\chi|\xi)$ for two values of $\beta$, at various values of $\xi=r/r_m$. Unlike $d=2$, here, we do not see any dramatic qualitative change in the angular distribution with increase in $\beta$. From simulations, we also identified the maximum $\xi_{\rm max}$ of $\Psi(\xi)$ and plotted it against $\beta$ (see Fig.~\ref{fig:rmax_3d}). For $\xi_{\rm max}>0$, the data was fitted to the expression 

\begin{equation}
     \xi_{\rm max}=\bigg(\frac{\beta}{\beta_c^{\rm sim}}-1\bigg)^a, \label{eq:rma}
\end{equation}
similar to Eq.~\ref{eq:rmax1}, from which we extracted the numerical values of the critical $\beta$ and the corresponding exponent:  $\beta_c^{\rm sim} \simeq $ 0.535  and $a\simeq 0.435$ in $d=3$. Note that in $d=3$, $\alpha=0$  for $\beta=1/2$ (see Eq.~\ref{eq:alpha}), which is close to the critical value found in simulations. This is similar to the corresponding observation in $d=2$, and very likely not entirely coincidental. However, at this stage, the generality of the above observation is only a conjecture and more mathematical work is required to establish it in $d>2$.

% and extracted the critical trap strength $\kappa_c\equiv D_r\zeta_t\beta_c$ (for fixed $D_r$) and the exponent $a$. The estimated values are $\kappa_c^{\rm sim}\simeq 0.5 \times 10^{-5}$ pN nm$^{-1}$ and $a\simeq 0.428$, obtained by plotting $\xi_{\rm max}$ on a log-scale against $\beta/\beta_c^{\rm sim}-1$. Here, $\beta_c^{\rm sim}$ is defined as the first  non-zero value of $\beta$ with non-zero $r_{\rm max}$, while the slope of the straight line fit gives $a$. The corresponding theoretical predictions are $\kappa_c\simeq 0.436 \times 10^{-5}$ pN nm$^{-1}$ (Eq.~\ref{eq:rmax} and Table~\ref{tab1}) and 0.5 (Eq.~\ref{eq:rmax}), respectively, corresponding to $\beta_c=1/3$. 

We conclude, therefore, that the positional distribution of the ABP undergoes a similar transition in $d=3$ as in $d=2$, but the critical value of $\beta$ is higher in $d=3$. As a word of caution, however, we would also like to point out that the multiplicative nature of the noise terms in Eq.~\ref{eq:3drotation} makes the forward Euler discretization scheme used here less reliable than in $d=2$. 

%\begin{comment}
\section{\label{sec:level4}Conclusions}
In this paper, we have presented a detailed analysis of the properties of the stationary state of an active Brownian particle (ABP) in a two-dimensional harmonic trap. We show that the relative significance of active motion vis-a-vis Brownian translational diffusion is controlled by a dimensionless diffusion coefficient $D_t^{\prime}$, which is the inverse of a P\'{e}clet number in the problem. When $D_t^{\prime}\gg 1$, the ABP effectively behaves like a passive Brownian particle, in equilibrium with the fluid. In this limit, the positional probability distribution of the particle asymptotically approaches the Boltzmann distribution, which is Gaussian. In the opposite limit, activity is dominant and the distribution acquires a strongly non-equilibrium character. The extreme limiting case where $D_t^{\prime}\to 0$ is especially interesting, and it is the situation that we addressed in detail in this paper. 

 In the limit $D_t^{\prime}\to 0$, our results strongly suggest that the positional probability distribution $\Psi(\xi)$ undergoes a continuous ``phase transition" between concave and convex shapes (around the origin) as the trap stiffness is increased and crosses a critical value. This is the central result in this paper. The explicit mathematical form of $\Psi(\xi)$ is non-Gaussian, even when it is concave, and is supported in a finite domain $r\in [0,r_m]$, where $r_m$ is an intrinsic length scale determined by the active velocity and trap stiffness. However, in the limit of very weak trap, the distribution assumes a Gaussian form, distinct from the equilibrium Boltzmann distribution, and characterized by an effective active temperature. 
 
The above remarkable results were derived from the exact Fokker-Planck equation for the particle's position, but under the assumption that the orientation angle of the self-propulsion velocity $\chi$ is a Gaussian variable. Although numerical simulations show that this assumption is violated in the super-critical (strong trap) regime, we find that results for the mean and variance of $\cos\chi$ do not deviate significantly from the predictions of the Gaussian theory. The critical value of the trap stiffness  and the corresponding critical exponent were determined from numerical simulations and found to agree closely with our theoretical predictions. Our numerical results also show that the angular distribution changes from unimodal to bimodal in the super-critical regime, a feature that had been observed in earlier experiments.  

We minimally extended our study to three dimensions, by simulating the dynamics of an ABP with $D_t=0$. We observed that the concave-convex transition in the positional distribution is present in $d=3$ also, with similar features as in $d=2$.The critical $\beta$ was found to be higher in $d=3$, but the exponent values were similar in both cases.
    
To conclude, our results presented in this paper provide new insights into the strongly non-equilibrium stationary states of an ABP in a harmonic trap. We hope that our study will stimulate further investigations in this area. In particular, we would be delighted to see more theoretical efforts towards a deeper understanding of the phase-transition like change in the particle's positional and angular  distributions in the strongly active limit, both in $d=2$ and $d=3$. It would also be interesting to explore the possible existence of such transitions in closely related models, e.g., run and tumble particle models. We have made preliminary progress in these directions, and hope to report our results in the near future. 

\begin{acknowledgments}
We thank our colleague Basudev Roy for introducing us to this interesting problem, sharing the experimental data and helpful discussions. We also gratefully acknowledge the computational facilities at P.G. Senapathy Centre for Computing Resources, IIT Madras. U.N acknowledges financial support by the Council for Scientific and Industrial Research (CSIR), India through a Junior Research Fellowship. We would also like to thank two anonymous reviewers for pointing out some mistakes in the previous versions of the manuscript and for many helpful suggestions.
\end{acknowledgments}

\appendix

\section{ Simplification of the term $\int \hat{\bf u}(\hat{\bf u}\cdot \nabla P)d^2\hat{\bf u}$}
To solve Eq.~\ref{eq:a1_main2}, we need to simplify the term

$\int \hat{\bf u}(\hat{\bf u}\cdot \nabla P)d^2\hat{\bf u}$. In plane polar coordinates,
\begin{equation}
    \nabla P=\frac{\partial P }{\partial r} \hat{\bf r}+\frac{1}{r}\frac{\partial P}{\partial \phi} \hat{\boldsymbol \phi} ;\quad  \hat{\bf u}=\cos{\chi} \hat{\bf r}+\sin{\chi} \hat{\boldsymbol{\phi}}. \label{eq:b1_main1}
\end{equation}
From azithmuthal symmetry, we conclude $P(r,\phi,\theta_u)\equiv P(r,\chi)$ where $\chi=\theta_u-\phi$, which also requires the variable transformation rules 
$\frac{\partial}{\partial \theta_u}\to \frac{\partial}{\partial \chi}$, $\frac{\partial}{\partial \phi}\to -\frac{\partial}{\partial \chi}$ for the derivatives. It follows that 
\begin{equation}
    \int \hat{\bf u}(\hat{\bf u}\cdot \nabla P) d^2\hat{\bf u}=\hat{\bf r}\int \cos{\chi} \left\{ \cos{\chi}\frac{\partial P}{\partial r}-\frac{\sin{\chi}}{r}\frac{\partial P}{\partial \chi} \right\} d\chi. \label{eq:b1_main3}
\end{equation}
Note that the $\hat{\boldsymbol \phi}$ component vanishes on account of periodic boundary conditions. Introduce the conditional distribution for $\chi$ by writing $P(r,\chi)$. After substituting this expression in Eq.~\ref{eq:b1_main3}, we find, for the radial component, 
\begin{equation}
\begin{aligned}
 \hat{\bf r}\cdot\int \hat{\bf u}(\hat{\bf u}\cdot \nabla P) d^2\hat{\bf u}=  \partial_r \left\{ \Phi(r) \langle \cos^2{\chi} \rangle \right\} +\frac{\Phi(r)}{r} \langle \cos{2\chi}\rangle. \label{eq:b1_main4}
 \end{aligned}
\end{equation}
In terms of the variance of $\cos\chi$, 
\begin{equation}
   \sigma^2 \equiv \sigma^2_{\cos{\chi}}=  \langle \cos^2{\chi} \rangle- \langle \cos{\chi}\rangle^2,  \label{eq:sigma_ax} 
  \end{equation}
and in terms of $\xi=r/r_m$, Eq.~\ref{eq:b1_main4} becomes 
\begin{equation}
\begin{aligned}
    \hat{\bf r}\cdot\int ~\hat{\bf u} (\hat{\bf u}\cdot \nabla P)~d^2\hat{\bf u}= \frac{1}{r_m^3}\bigg(\sigma^2 +\xi^2 \bigg) \partial_{\xi} \Psi(\xi)&\\ +\frac{\Psi(\xi)}{r_m^3}\bigg( \partial_{\xi} \sigma^2 +4\xi+\frac{2 \sigma^2-1}{\xi}\bigg)&\label{eq:b1_main5}
    \end{aligned}
\end{equation}
\section{$\sigma_{\cos{\chi}}^2$ when $\chi$ is assumed Gaussian}
For a Gaussian variable $\chi$ with probability distribution  $p(\chi)=\frac{1}{\sqrt{2\pi}\sigma^2}e^{-\frac{\chi^2}{2\sigma^2}}$, 
$\langle e^{iq\chi} \rangle = e^{-\frac{q^2\sigma^2}{2}}$. It follows that, for integer $n$, 
\begin{equation}
    \langle \cos{n\chi}\rangle = \langle \cos{\chi}\rangle^{n^2}
    \label{eq:cosn}
\end{equation}
Hence, 
\begin{equation}
\begin{aligned}
  \langle \cos^2{\chi} \rangle =&\frac{1}{2}[\langle \cos 2\chi \rangle+1]
  \equiv \frac{1}{2}(1+\langle \cos{\chi}\rangle^4)
  \end{aligned}
\end{equation}
It follows that the variance of $\cos\chi$ is given by 
\begin{equation}
    \sigma^2_{\cos{\chi}}=\frac{1}{2}\left( 1-\langle \cos{\chi}\rangle^2\right)^2 \label{eq:app_var}
\end{equation}
Eq.~\ref{eq:app_var} when combined with Eq.~\ref{eq:coschi} leads to 
Eq.~\ref{eq:variance_cos}.

\section{The general equation for $g(\xi)$}
Consider the stationary state equation (Eq.~\ref{eq:stationaryFPE}). Upon multiplying by $\hat{\bf u}$ and integrating over the same, we find 
\begin{equation}
\begin{aligned}
   & D_t \int \hat{\bf u} \nabla^2 P d^2\hat{\bf u}+D_r\int \hat{\bf u} \nabla^2_{\hat{\bf u}}P d^2\hat{\bf u}+2k \int \hat{\bf u} P d^2\hat{\bf u}\\&-u_0\int \hat{\bf u} (\hat{\bf u}\cdot \nabla P) d^2\hat{\bf u}+k \int \hat{\bf u} ({\bf r}\cdot \nabla P) d^2\hat{\bf u}=0
\end{aligned}
\label{eq:C2}
\end{equation}
Using Eq.~\ref{eq:gr}, Eq.~\ref{eq:dimensionless} and Eq.~\ref{eq:sigma_ax}, the following relations are easily proved in $d=2$. Here, we define $\sigma^2\equiv \sigma_{\cos\chi}^2$.
\begin{eqnarray}
     \int \hat{\bf u} (\hat{\bf u}\cdot \nabla P) d^2\hat{\bf u}=\frac{\hat{\bf r}}{r_m^3}\Bigg[(\sigma^2 +g^2)\partial_{\xi}\Psi+\nonumber\\
   +\Psi\Big( \frac{2\sigma^2+2g^2-1}{\xi}+\partial_{\xi}\sigma^2+2g\partial_{\xi}g \Big) \Bigg],  
     \label{eq:C4}
\end{eqnarray}
which generalizes Eq.~\ref{eq:b1_main5} in Appendix A. Next, 
\begin{equation}
\int \hat{\bf u}(\mathbf{r}\cdot \nabla P)d^2\hat{\bf u}=
   \frac{\xi}{r_m^2}\partial_{\xi}(g \Psi )\hat{\bf r}, 
   \label{eq:C5}
\end{equation}
\begin{equation}
     \int \hat{\bf u} \nabla^2 P  d^2\hat{\bf u}=\frac{\hat{\bf r}}{r_m^4}\Bigg[ \partial_{\xi}^2 (g\Psi) +\frac{1}{\xi} \partial_{\xi}(g\Psi)-\frac{g\Psi}{\xi^2} \Bigg], 
     \label{eq:C6}
\end{equation}
and, finally, 
\begin{equation}
   \int \hat{\bf u} \nabla^2_{\hat{\bf u}} P  d^2\hat{\bf u}=-\frac{\Psi g}{r_m^2} ~\hat{\bf r}\label{eq:C7}
\end{equation}
Using Eq.~\ref{eq:C4}, Eq.~\ref{eq:C5}, Eq.~\ref{eq:C6} and Eq.~\ref{eq:C7}, we rewrite Eq.~\ref{eq:C2} as 
\begin{eqnarray}
D_t^{\prime}\left(\partial^2_{\xi} g+\frac{\partial_{\xi}g}{\xi}-\frac{g}{\xi^2}\right)-\partial_{\xi}\sigma^2+g\left(\frac{g}{\xi}-\frac{1}{\beta}\right)\nonumber\\-(g-\xi)\partial_{\xi}g-\frac{2\sigma^2+2g^2-1}{\xi}-\frac{\sigma^2}{D_t^{\prime}}\left(g-\xi \right)=0~~~~
\label{eq:fequation}
\end{eqnarray}

where $D_t^{\prime}$ and $\beta$ are given by Eq.~\ref{eq:dimensionlessparameter} and Eq.~\ref{eq:beta}, respectively. In the limit $D_t^{\prime}\to 0$, Eq.~\ref{eq:fequation} reduces to 
\begin{equation}
   \sigma^2 (g-\xi)=0, 
\end{equation}
which is consistent with Eq.~\ref{eq:gr1}, since $\sigma^2>0$. 
In the vicinity of $\xi=0$, let us expand the function $g(\xi)$ in a power-series: $g(\xi)\sim g^{\prime}(0) \xi+(g^{\prime\prime}(0)/2)\xi^2+{\mathcal O}(\xi^3)$. Similarly, the variance of $\cos\chi$ may be expanded as  $\sigma^2_{\cos\chi}(\xi)\sim \sigma^2_{\cos\chi}(0)+{\mathcal O}(\xi)$. Upon substituting these expansions in Eq.~\ref{eq:fequation}, we arrive at the following useful relation: 
\begin{equation}
    g^{\prime}(0)=\frac{2\sigma^2_{\cos\chi}(0)-1}{D_t^{\prime}}.
    \label{eq:fprime1}
\end{equation}

\section{$\sigma^2_{\cos\chi}(0)=\frac{1}{2}$ for $D_t^{\prime}\geq 0$}
In the vicinity of $r=0$, Eq.~\ref{eq:stationaryFPE} assumes the following form, in terms of $r$ and $\chi$: 
\begin{equation}
\begin{aligned}
 \bigg(D_r +\frac{D_t}{r^2}\bigg)\frac{\partial^2 P}{\partial \chi^2}+\bigg( \frac{D_t}{r}- u_0 \cos{\chi}\bigg)\frac{\partial P}{\partial r}&\\ +\frac{u_0 \sin{\chi}}{r}\frac{\partial P}{\partial \chi}+D_t \frac{\partial^2 P}{\partial r^2}=0&
    \end{aligned} \label{eq:d3}
\end{equation}
The singular terms in the above equation can be separated by subjecting $P(r,\chi)$ to the Taylor expansion
\begin{equation}
    P(r,\chi)\sim \sum_{n=0}^{\infty} a_n(\chi)r^n
    \label{eq:Taylor}
\end{equation}
in the neighbourhood of $r=0$, where $a_n(\chi)$ are periodic functions of $\chi$. Eq.~\ref{eq:Taylor} when substituted in Eq.~\ref{eq:d3} allows the functions $a_n(\chi)$ to be determined, by separating terms proportional to different powers of $1/r$. The first three equations are 
\begin{eqnarray}
\begin{aligned}
& \mathcal{O}(r^{-2}):&~~~~a_0^{\prime\prime}(\chi)=0\\
 & \mathcal{O}(r^{-1}):&~~~~D_t a_1(\chi)+ u_0\sin{\chi} a_0^{\prime}(\chi)=0\\
 & \mathcal{O}(r^0):&~~~~~a_2^{\prime\prime}(\chi)+4 a_2(\chi)=0. 
    \end{aligned}
    \label{eq:dX}
\end{eqnarray}
The first equation in Eq.~\ref{eq:dX} implies that $a_0^{\prime}(\chi)=C$, a constant, but periodicity requires that only $C=0$ is possible. From the second equation, it thus follows that $a_1(\chi)=0$ while the last one shows that $a_2(\chi)$ is a periodic function of $\chi$, as it should be. From Eq.~\ref{eq:r9}, and using Eq.~\ref{eq:Taylor}, we find $f(\chi|0)\propto a_0(\chi)$, and hence $f(\chi|0)=1/2\pi$ for $-\pi\leq \chi\leq \pi$. It follows that $\sigma^2_{\cos\chi}(0)=1/2$ in general.

\end{document}